\documentclass[structabstract]{aa}  
                                   
\usepackage{graphicx}
\usepackage{lscape}
\usepackage{amsmath}
\usepackage{txfonts}
\usepackage{url}
\usepackage{natbib}
\usepackage{xcolor}
\usepackage{color}
\usepackage{multirow}

\begin{document}
   \title{Grids of stellar models with rotation}

   \subtitle{II. WR populations and supernovae/GRB progenitors at \textit{Z} = 0.014}

   \author{C. Georgy\inst{1},
        S. Ekstr\"om\inst{2},
        G. Meynet\inst{2},
        P. Massey\inst{3},
        E. M. Levesque\inst{4},
        R. Hirschi\inst{5,6},
        P. Eggenberger\inst{2},
        A. Maeder\inst{2}
}

   \authorrunning{Georgy et al.}

   \institute{Centre de Recherche Astrophysique de Lyon, Ecole Normale Sup\'erieure de Lyon, 46, all\'ee d'Italie, F-69384 Lyon cedex 07, France
          \and Geneva Observatory, University of Geneva, Maillettes 51, CH-1290 Sauverny, Switzerland
          \and Lowell Observatory, 1400 W Mars Hill Road, Flagstaff, AZ 86001, USA
          \and CASA, Department of Astrophysical and Planetary Sciences, University of Colorado 389-UCB, Boulder, CO 80309, USA
          \and Astrophysics group, EPSAM, Keele University, Lennard-Jones Labs, Keele, ST5 5BG, UK
          \and Institute for the Physics and Mathematics of the Universe, University of Tokyo, 5-1-5 Kashiwanoha, Kashiwa, 277-8583, Japan}

   \date{Received ; accepted }

 
\abstract
{In recent years, many very interesting observations have appeared concerning the positions of Wolf-Rayet (WR) stars in the Hertzsprung-Russell diagram (HRD), the number ratios of WR stars, the nature of type Ibc supernova (SN) progenitors, long and soft gamma ray bursts (LGRB), and the frequency of these various types of explosive events. These observations represent key constraints on massive star evolution.} 
{We study, in the framework of the single-star evolutionary scenario, how rotation modifies the evolution of a given initial mass star towards the WR phase and how it impacts the rates of type Ibc SNe. We also discuss the initial conditions required to obtain collapsars and  LGRB.}
{We used a recent grid of stellar models computed with and without rotation to make predictions concerning the WR populations and the frequency of different types of core-collapse SNe. Current rotating models were checked to provide good fits to the following features: solar luminosity and radius at the solar age, main-sequence width, red-giant and red-supergiant (RSG) positions in the HRD, surface abundances, and rotational velocities.}
{Rotating stellar models predict that about half of the observed WR stars and at least half of the type Ibc SNe may be produced through the single-star evolution channel. Rotation increases the duration of the WNL and WNC phases, while reducing those of the WNE and WC phases, as was already shown in previous works. Rotation increases the frequency of type Ic SNe. The upper mass limit for type II-P SNe is $\sim 19.0\, M_{\sun}$ for the non rotating models and $\sim 16.8\, M_{\sun}$ for the rotating ones. Both values agree with observations. Moreover, present rotating models provide a very good fit to the progenitor of SN 2008ax. We discuss future directions of research for further improving the agreement between the models and the observations. We conclude that the mass-loss rates in the WNL and RSG phases are probably underestimated at present. We show that up to an initial mass of $40\, M_{\sun}$, a surface magnetic field inferior to about $200\,\mathrm{G}$ may be sufficient to produce some braking. Much lower values are needed at the red supergiant stage. We suggest that the presence/absence of any magnetic braking effect may play a key role in questions regarding rotation rates of young pulsars and the evolution leading to LGRBs.}
{}

\maketitle
\section{Introduction}

Wolf-Rayet (WR) stars are characterised by four important observed features \citep[see the recent reviews by][]{Massey2003b, Crowther2007a}:
\begin{enumerate}
\item They are associated with young massive star regions.
\item They present broad and strong emission lines.
\item They are hot ($\log (T_\text{eff}/\text{K}) \gtrsim 4$) and luminous stars ($\log (L/L_{\sun})> 5.0$).
\item The chemical composition of their surface shows signs of H-burning (WN-type) and/or He-burning processes (WNC, WC, WO-type).
\end{enumerate}

These features can be satisfactorily explained if WR are massive evolved stars whose surface composition has been changed by mass loss and/or internal mixing. The mass loss can be due either to stellar winds and/or to the loss of the envelope through a Roche lobe overflow (RLOF) in a close-binary system.

The estimated total number of WR stars in the Milky Way is about 6000--6500 \citep{Shara2009a}. It means that about 1 out of 20 million stars is a WR star in the Galaxy. To date, 476 galactic WR stars have been detected, \textit{i.e.} about 7--8\% of the total Milky Way population \citep{Mauerhan2011a}. Although these stars are quite rare, they are important in astrophysics for many reasons:
\begin{itemize}
\item They represent an evolved state of the most massive stars \citep[see \textit{e.g.}][]{ Schnurr2008a,Rauw2004a,Lamers1991a}.
\item They allow us to check the nuclear reaction chains during the H- and He-burning phases \citep[see \textit{e.g.}][]{Dessart2000a}.
\item They suffer strong mass loss, making them very interesting laboratories for studying the physics of stellar winds. Many of them are surrounded by extended nebulae \citep{Stock2010a}.
\item They are candidate progenitors for type Ibc supernovae (SNe)\footnote{\footnotesize{In this paper, we call the sample composed of type Ib and type Ic SNe type Ibc SNe.}}, which give birth to either a neutron star (NS) or a black hole \citep[BH, ][]{Smartt2009b}.
\item They may be the progenitors of long and soft gamma-ray bursts (LGRBs), or at least of part of them \citep{Woosley1993a,Woosley2011a}.
\item They are important contributors to the injection of new synthesised nuclear species in the interstellar medium through their winds and possibly also through their SN ejecta, making them an important agent of chemical evolution in galaxies. In particular, they may be significant sources of $^{12}$C \citep{Maeder1992a} , $^{19}$F \citep{Meynet2000b} and $^{26}$Al \citep{Dearborn1984a, Palacios2005a}, to cite only a few elements. A WR star probably injected the $^{26}$Al present at the birth of the solar system \citep{Tatischeff2010a,Montmerle2010a,Gounelle2011a}.
\item The presence of WR stars can be detected through the analysis of the integrated light spectrum of very distant galaxies \citep{Allen1976a,Conti1991a}. Indeed, their broad emission lines can be observed superposed to the galactic continuum in young active star-forming systems, making these stars very useful tracers of distant starburst regions.
\end{itemize}
Amongst the questions concerning the WR stars, which are a source of lively debate these days, we cite the following two:
\begin{itemize}
\item In what way are the single-star and close-binary channels significant for explaining the observed WR populations  \citep{Podsiadlowski1992a,Izzard2004a,Eldridge2008a}? Do these two scenarios have different effects at various metallicities? Depending on the answers to these questions, the age associated with a WR population might be quite different, the range of initial masses evolving into the WR phase being likely different in the single and binary scenario.
\item What kind of SNe do the WR stars produce? If they collapse into a BH, is the latter associated with a faint or does it prevent any SN event? Do they give birth to a type Ib or type Ic SN? Do WR stars produced through the single and binary scenario have the same SN output? Answers to these questions are important for the contribution of these stars to the chemical evolution of the galaxies, for interpreting the observed frequencies of type Ibc SNe, and to check the possibility that these stars are in some circumstances associated to LGRB events, since in a handful of cases the spectrum of a type Ic SN has been observed in association with an LGRB \citep[][and references therein]{Galama1998a,Chornock2010a,Berger2011a}.
\end{itemize}

In the present work, we aim to make progress towards providing answers to the above questions, analysing the results we have obtained in our most recent solar-metallicity grid of stellar models \citep[][hereafter Paper~I]{Ekstrom2012a}\defcitealias{Ekstrom2012a}{Paper~I}. These models, which include the effects of rotation, are able to reproduce many observed characteristics: 
\begin{itemize}
\item the characteristics of the Sun at its present age,
\item the observed width of the main-sequence (MS) band in the Hertzsprung-Russell diagram (HRD),
\item the positions of red giants and red supergiants (RSG) in the HRD,
\item the observed surface composition changes in B-type dwarfs and supergiants,
\item the observed averaged rotational surface velocities.
\end{itemize}
We aim to check whether these models are also able to reproduce the observed ratios of WR to O-type stars, of WN to WC stars, and of WNC stars (a transition stage between the WN and WC stages where both H- and He-burning products are observed)  to WR stars at solar metallicity. This study will fuel the discussion of how rotation modifies the evolutionary scenarios in the upper part of the HRD. We also study the nature of the SNe arising from stars more massive than about $8\, M_{\sun}$.

The paper is structured as follows: Sect.~\ref{SecPhymod} briefly recalls the main physical ingredients of the models. The WR stellar models are presented in Sect.~\ref{SecWRmodels}. Comparisons with the observations are performed in Sect.~\ref{SecPopu}. The nature of the progenitors of type Ibc  SNe is the subject of Sect.~\ref{SecSNe}, while Sect.~\ref{SecGRB} focuses on the questions of the rotation rate of pulsars and the conditions for obtaining collapsars and LGRB. A synthesis of the main results is presented in Sect.~\ref{SecConclu}.

\section{Physical ingredients of the models \label{SecPhymod}}

The physical ingredients of the models are described in detail in \citetalias{Ekstrom2012a}. We recall the main features here:
\begin{itemize}
\item The initial abundances were set to $X=0.720$, $Y=0.266$ and $Z=0.014$. The mixture of heavy elements is taken to be the same as in \citet{Asplund2005a} except for the Ne abundance, which we took from \citet{Cunha2006a}.
\item The opacities and nuclear reaction rates were updated \citepalias[see details in][]{Ekstrom2012a}.
\item The convective core is extended with an overshoot parameter $d_\text{over}/H_P=0.10$ starting from the Schwarzschild limit.
\item The outer convective zone is treated according to mixing length theory, with  $\alpha_\text{MLT}=1.6$. The most luminous models ($M\geq40\, M_{\sun}$) can have a density inversion near the surface due to supra-Eddington luminosity layers. In those models, the mixing length is based on the density scale: $\alpha_\text{MLT}=\ell/H_\rho = \ell (\alpha - \delta \nabla) / H_P = 1$. The turbulence pressure and the energy turbulent flux are included \citep[see][Sect.~5.5]{Maeder2009a}.
\item Models with rotation account for the effects of the strong horizontal turbulence, the vertical shear, and the meridional circulation as explained in \citetalias{Ekstrom2012a}. No magnetic field is assumed.
\item A treatment allowing the precise conservation of angular momentum in rotating models was implemented.
\item  The radiative mass-loss rate adopted is from \citet{Vink2001a}. In the domains not covered by this prescription, we used that of \citet{deJager1988a}. For red (super)giants, we used the \citet{Reimers1975a,Reimers1977a} formula for stars up to $12\, M_{\sun}$, with $\eta=0.5$, and that of \citet{deJager1988a} from $15\, M_{\sun}$ and up for stars with $\log (T_\text{eff}/\text{K}) > 3.7$, or a linear fit on the data from \citet{Sylvester1998a} and \citet{vanLoon1999a} \citep[see][]{Crowther2001a} for $\log (T_\text{eff}/\text{K}) < 3.7$. The WR stars are computed with the \citet{Nugis2000a} prescription, or the \citet{Grafener2008a} recipe in the restricted validity domain of this prescription. Note that it may occur that the WR mass-loss rate is lower than the one that would result from the \citet{Vink2001a} prescription. In that case we keep to the \citet{Vink2001a} prescription until the WR prescription becomes higher. These mass-loss rates account for some clumping effects \citep{Muijres2011a} and are a factor of 2--3 lower than the rates used in the standard 1992 grid.
\item When, for massive stars ($> 15\, M_{\sun}$) in the RSG phase, some points in the most external layers of the stellar envelope exceed the Eddington luminosity of the star ($L_\text{Edd} = 4\pi cGM/\kappa$, with $\kappa$ being the opacity), we artificially increase the mass-loss rate of the star (computed according to the prescription described above) by a factor of $3$ so that the time-averaged mass-loss rate during the RSG phase for 20 and $25\, M_{\sun}$ stars is of the order of the mass-loss rates estimated by \citet{vanLoon2005a} for RSGs. 
\end{itemize}

\section{The WR stellar models}\label{SecWRmodels}

\begin{table*}
\caption{O-type star, RSG, and WR lifetimes for the most massive models.}
\label{TabWRtau}
\centering
\begin{tabular}{r l| l l l l l l l l}
\hline\hline
    $M_\text{ini}$ & v$_\text{ini}/\text{v}_\text{crit}$ & $\tau_\text{RSG}$ & $\tau_\text{O-star}$ & $\tau_\text{WR}$ & $\tau_\text{WNL}$ & $\tau_\text{WNE}$ & $\tau_\text{WNC}$ & $\tau_\text{WC}$ & $\tau_\text{WO}$    \\
\hline
   \rule[0mm]{0mm}{3mm} $120$  & $0$      & -- &  $2.15\cdot 10^{6}$  & $3.97\cdot 10^{5}$  & $1.15\cdot 10^{5}$  & $9.28\cdot 10^{3}$  & $2.66\cdot 10^{2}$  & $2.72\cdot 10^{5}$  &   --   \\
                 &  $0.4$   & -- & $ 2.34\cdot 10^{6}$  & $1.22\cdot 10^{6}$  & $8.88\cdot 10^{5}$  & $2.46\cdot 10^{4}$  & $3.92\cdot 10^{3}$  & $3.06\cdot 10^{5}$  &   --   \\
     $85$   &  $0$       & -- & $2.41\cdot 10^{6}$  & $3.52\cdot 10^{5}$  & $3.89\cdot 10^{4}$  & $2.08\cdot 10^{4}$  & $1.63\cdot 10^{3}$  & $2.92\cdot 10^{5}$  &   --   \\
                 &  $0.4$    & -- &  $2.87\cdot 10^{6}$  & $1.19\cdot 10^{6}$  & $8.90\cdot 10^{5}$  & $1.11\cdot 10^{4}$  & $5.38\cdot 10^{3}$  & $2.91\cdot 10^{5}$  &   --   \\
     $60$   &  $0$       & -- &  $3.00\cdot 10^{6}$  & $3.97\cdot 10^{5}$  & $6.13\cdot 10^{4}$  & $5.15\cdot 10^{4}$  & $2.20\cdot 10^{3}$  & $2.85\cdot 10^{5}$  &   --   \\
                 &  $0.4$   & -- &  $3.95\cdot 10^{6}$  & $9.05\cdot 10^{5}$  & $6.12\cdot 10^{5}$  & $3.03\cdot 10^{4}$  & $3.20\cdot 10^{4}$  & $2.63\cdot 10^{5}$  &   --   \\
     $40$   & $0$       & $6.00\cdot 10^{4}$ &  $3.88\cdot 10^{6}$  & $7.46\cdot 10^{4}$  & $3.59\cdot 10^{4}$  & $3.88\cdot 10^{4}$  &   --    &   --    &   --   \\
                 &  $0.4$   & -- &  $5.72\cdot 10^{6}$  & $4.10\cdot 10^{5}$  & $2.18\cdot 10^{5}$  & $9.59\cdot 10^{3}$  & $2.30\cdot 10^{4}$  & $1.82\cdot 10^{5}$  &   --   \\
     $32$   &  $0$       & $1.17\cdot 10^{5}$ &  $4.56\cdot 10^{6}$  & $6.49\cdot 10^{3}$  & $6.07\cdot 10^{3}$  & $4.13\cdot 10^{2}$  &   --    &   --    &   --   \\
                 &  $0.4$   & -- &  $6.37\cdot 10^{6}$  & $4.54\cdot 10^{5}$  & $2.77\cdot 10^{5}$  & $1.99\cdot 10^{4}$  & $2.59\cdot 10^{4}$  & $1.57\cdot 10^{5}$  &   --   \\
     $25$   &  $0$       & $2.96\cdot 10^{5}$ &  $5.29\cdot 10^{6}$  & $1.01\cdot 10^{3}$  & $1.01\cdot 10^{3}$  &   --    &   --    &   --    &   --   \\
                 &  $0.4$   & $1.30\cdot 10^{5}$ &  $6.56\cdot 10^{6}$  & $5.80\cdot 10^{4}$  & $5.80\cdot 10^{4}$  &   --    &   --    &   --    &   --   \\
     $20$   &  $0$       & $5.49\cdot 10^{5}$ &  $5.82\cdot 10^{6}$  &   --    &   --    &   --    &   --    &   --    &   --   \\
                 &  $0.4$   & $3.45\cdot 10^{5}$ & $6.50\cdot 10^{6}$ & $1.80\cdot 10^{3}$ & $1.80\cdot 10^{3}$  &   --    &   --    &   --    &   --   \\
     $15$   &  $0$       & $8.79\cdot 10^{5}$ &  $2.23\cdot 10^{5}$  &   --    &   --    &   --    &   --    &   --    &   --   \\
                 &  $0.4$   & $1.17\cdot 10^{6}$ &    --    &   --    &   --    &   --    &   --    &   --    &   --   \\
\hline
\end{tabular}
\tablefoot{The initial masses are given in $M_{\sun}$. Lifetimes are given in years.\\
}
\end{table*}

\subsection{HR diagram}

\begin{figure*}
\centering
\includegraphics[width=.45\textwidth]{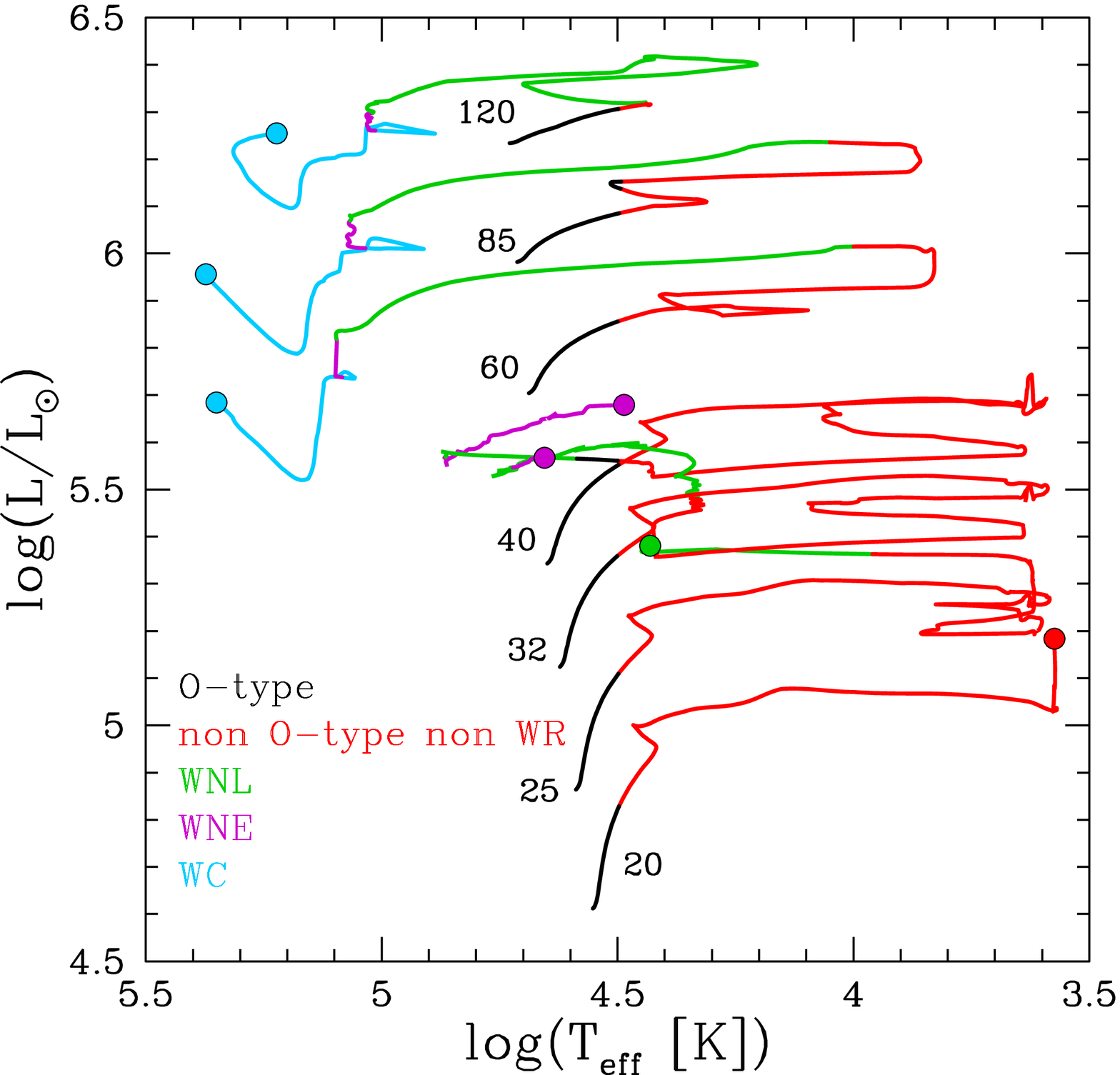}\hspace{.5cm}\includegraphics[width=.45\textwidth]{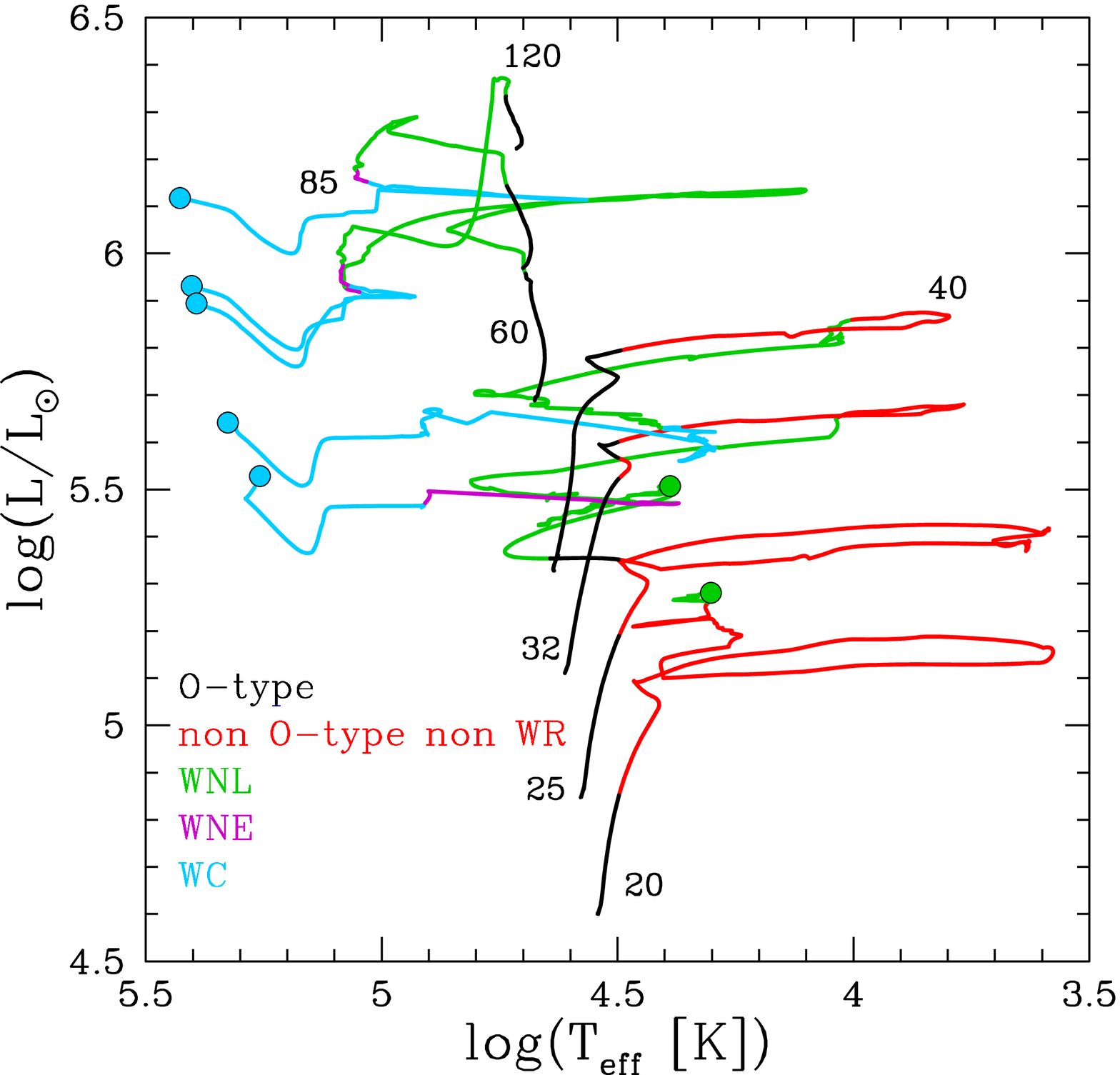}
  \caption{HRD of the massive models from 20 to $120\, M_{\sun}$ with the different types/phases marked in colours (O-type: blue; neither O-type nor WR: red; WNL: green; WNE: purple; WC: cyan). \textit{Left:} Non-rotating models. \textit{Right:} Rotating models. We plotted the effective temperature at the surface of the hydrostatic core. The endpoints of the tracks are indicated by a circle.}
     \label{FigHRDWR}
\end{figure*}

We used the following criteria to determine the type of the star at a given time \citep{Meynet2003a,Smith1991a}:
\begin{itemize}
\item A star with $\log(T_\text{eff}/\text{K}) > 4.0$ and a surface hydrogen mass fraction $X_\text{H} < 0.3$ is considered as a WR star\footnote{\footnotesize{The duration of the corresponding WR phase is not very sensitive to the choice of this limit if it remains in the range $0.3$ -- $0.4$ as mentioned by \citet{Meynet2003a}.}}.
\item A star with $\log(T_\text{eff}/\text{K}) > 4.5$ that is not a WR star is an O-type star.
\item A WR star with a surface hydrogen mass fraction $X_\text{H} > 10^{-5}$ is a WNL star.
\item A WR star without hydrogen and with a carbon surface abundance lower than the nitrogen abundance is a WNE star.
\item A WR star without hydrogen and with a carbon surface abundance higher than the nitrogen abundance is a WC or a WO star. If the surface ratio $\frac{\text{C} + \text{O}}{\text{He}}$ (in number) is less than 1, the star is classified as a WC; otherwise, it is classified as a WO.
\item A WR star without hydrogen, and with a surface ratio $\frac{X_\text{C}}{X_\text{N}}$ between  $0.1$ and $10$ is a WNC star.
\end{itemize}
The O-type, WNL, WNE, WC, and WO phases are exclusive. The WNC phase lies between the WNE and WC phases. We point out here that the observational classification of WR stars is based on spectroscopic characteristics and not on surface abundances. A firm classification of our models would therefore require computing the output spectrum. While some attempts in that direction have already been made \citep{Schaerer1996a,Schaerer1996b,Schaerer1997a}, this is by far not yet a standard procedure. We decided here to follow the usual procedure most often found in the literature to connect WR spectral types to surface abundances. Below we present some numerical examples that illustrate how numbers change when different rules are adopted for the connection between surface abundances and WR subtypes. In Fig.~\ref{FigHRDWR} we show where each WR phase occurs in the HRD for the models between $20$ and $120\, M_{\sun}$. 

Comparing the rotating and non-rotating models, we note the following differences:
\begin{itemize}
\item The MS rotating tracks for the most massive stars ($M \ge 60\, M_{\sun}$ and $v_{\rm ini} \ge 350$ km s$^{-1}$) follow a nearly homogeneous evolution during the MS phase \citep{Maeder1987a}. As a numerical example, the mass fraction of hydrogen at the stellar surface is  0.32 in the rotating $120\, M_{\sun}$ model when the mass fraction of hydrogen at the centre is 0.28. We note that although these models remain quite compact and hence blue during the MS, their surface velocity decreases a lot. Indeed, at the end of the core H-burning phase, the surface velocity is only a few km s$^{-1}$ for stars with initial masses equal or superior to $60\, M_{\sun}$. This is caused by the high mass-loss rates undergone by these stars. 
\item Without rotation, only the $120\, M_{\sun}$ enters the WR phase at the end of the MS, while with rotation, the $60\, M_{\sun}$ and more massive stars already enter into the WR phase during the MS phase. 
\item A small range of initial masses go through an RSG stage before entering the WR regime. These models reach  $\log (T_\text{eff}/\text{K}) \sim 3.5$ at the RSG stage, and then, due to the strong mass loss, evolve back to the blue part of the HRD. They finally become WR stars at the very end of their evolution, ending as a WNL or WNE stars with an effective temperature (at the border of the hydrostatic core) of roughly $\log (T_\text{eff}/\text{K}) \sim 4.5$. Rotation restricts the mass domain of the family of tracks evolving into a WR phase after a RSG phase (see Fig.~\ref{FigWRtau}). As a consequence, rotation decreases the maximum luminosity of the RSGs from 5.7 without rotation to about 5.4 with rotation. The last value agrees better with the upper limit of RSGs, which was found by \citet{Levesque2005a} to be between 5.2 and 5.4.
\item The minimum luminosity reached by the WN and WC stars is given by the minimum initial masses of the stars that go through WNL and WC phases (see Table 2). They are lowered by rotation: for non-rotating models, the minimum luminosity of WNL and WC stars is 5.37 and 5.5, respectively, while for the rotating models it is equal to 5.25 and 5.35, respectively, {\it i.e.} in the same range as the maximum luminosity for RSG stars. 
\item Both rotating and non-rotating models predict that the least luminous WR stars are WNL stars. The least luminous stars are produced by the least massive star that just succeeds to enter the WNL phase at the end of its evolution. Note that if the mass-loss rates are allowed to be higher than accounted for in the present grid, there might be situations where the least luminous stars would be WC stars originating from stars well above the minimum initial mass of stars evolving into a WR stage. This is what was obtained in the grid with enhanced mass loss by \citet{Maeder1994a}, for instance. The nature of the less luminous WR stars therefore depends quite closely on the mass-loss rate history. 
\end{itemize}

\subsection{WR lifetimes and mass limits \label{SecTauWR}}

\begin{figure*}
\centering
\includegraphics[width=.45\textwidth]{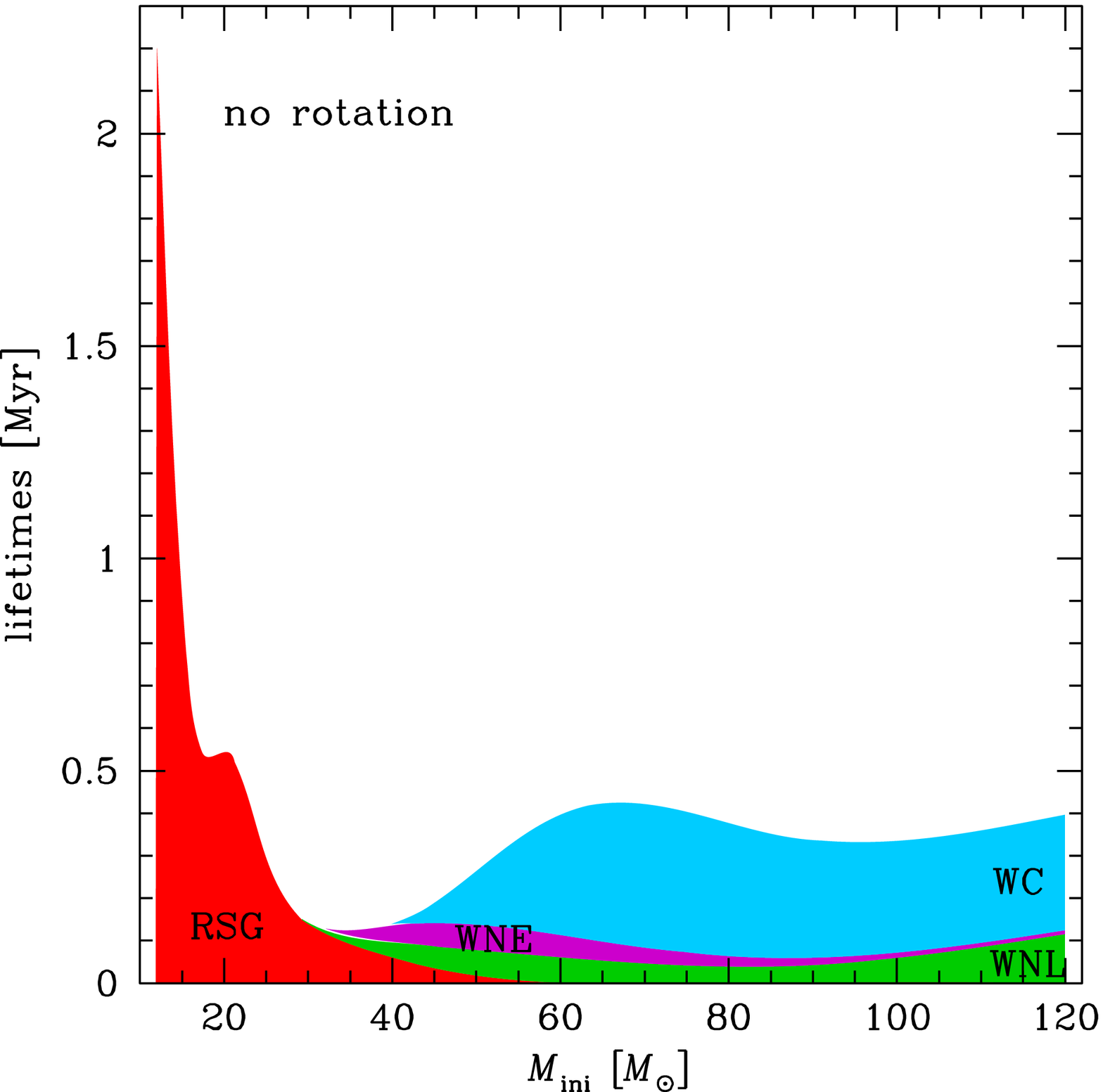}\hspace{.5cm}\includegraphics[width=.45\textwidth]{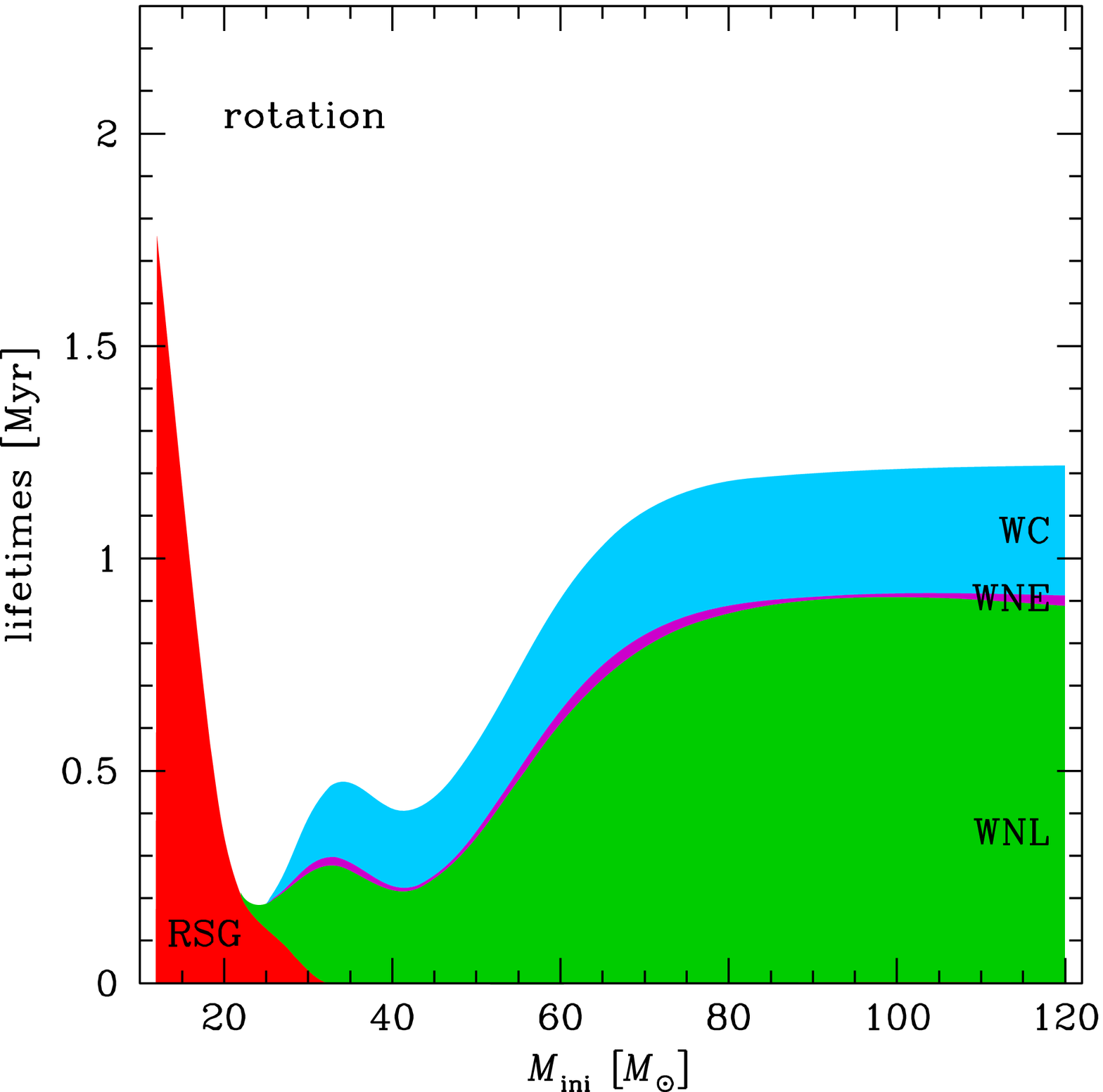}
  \caption{Lifetimes in the RSG phase \citep[defined as stars with $\log (T_\text{eff}/\text{K}) < 3.66$, see][]{Eldridge2008a} and in the different phases of WR stars. \textit{Left:} Non-rotating models. \textit{Right:} Rotating models.}
     \label{FigWRtau}
\end{figure*}

The WR lifetimes are indicated in Table \ref{TabWRtau} and are plotted as a function of the initial mass in Fig.~\ref{FigWRtau}. The minimum masses leading to a WR star (or a subtype) $M_\text{min}^\text{WR}$  are given in Table~\ref{TabMlim}. The mass limits are determined as in \citet{Georgy2009a}.

Qualitatively, the results do not differ from those presented in previous papers \citep{Meynet2003a,Meynet2005a}. The general trend is an increase of the WR phase lifetime with increasing mass and rotation. The most spectacular increase concerns the duration of the WNL phase for rotating models, which is several times longer than for non-rotating ones. On the other hand, the WNE phase almost disappears for rotating models. Again, this is a consequence of rotational mixing. To understand this, we have to keep in mind a few specific points. First, a WNE phase implies some pure He layers, which obviously appear only when H-burning is terminated in those layers. Second,  the width in mass of the pure He-rich region, which when uncovered gives the WNE phase, is decreasing rapidly with time because of the fast growing He-burning core that enriches the central layers in carbon and oxygen. Third, mixing prolongs the WNL stage, as just seen above, and consequently leaves more time for the star to evolve into the core He-burning phase, which in turn reduces the mass of pure He-layers. This makes the WNE phase very short in rotating models, or even more generally in any models with some efficient internal mixing.


\begin{table*}
\caption{Mass ranges for the various WR- and SN-remnant-types (in $M_{\sun}$) deduced from our models, and comparison with previous works.}
\label{TabMlim}
\centering
\begin{tabular}{c r  r| c c c c}
\hline\hline
\rule[0mm]{0mm}{3mm} & &    & O-type & WNL & WNE & WC \\
\cline{2-7}
\rule[0mm]{0mm}{3mm} & \multicolumn{2}{l|}{This work} &  &  &  & \\
    & & rot. & 15.8 -- 20.0 & 20.0 -- 25.3 & 25.3 -- 27.0 & 27.0 -- 120.0 \\
    & & no rot. & 15.0 -- 25.0 & 25.0 -- 31.7 & 31.7 -- 40.5 & 40.5 -- 120.0 \\
& \multicolumn{2}{l|}{This work (High WNE\tablefootmark{1})} &  &  &  & \\
    & & rot. & 15.8 -- 20.0 & 20.0 -- 21.6 & 21.6 -- 27.0 & 27.0 -- 120.0 \\
    & & no rot. & 15.0 -- 25.0 & 25.0 -- 29.2 & 29.2 -- 40.5 & 40.5 -- 120.0 \\
& \multicolumn{2}{l|}{\citet{Georgy2009a}} &  &  &  & \\
    & & rot. & & 23.0 -- 26.0 & 26.0 -- 29.0 & 29.0 -- 120.0 \\
\cline{2-7}
\rule[0mm]{0mm}{3mm} & &    & SN II-P & SN II-L/n & SN Ib & SN Ic \\
\cline{2-7}
\multirow{12}{0.5cm}{\rotatebox[origin=c]{90}{BH $\rightarrow$ bright SN\tablefootmark{3}}$\left\{\rule{0cm}{2.15cm}\right.$} & \multicolumn{2}{l|}{\rule[0mm]{0mm}{3mm}This work (low SN Ic\tablefootmark{2})} &  &  &  & \\
    & & rot. & 8.0 -- 16.8 & 16.8 -- 25.0 & 25.0 --  31.1 & 31.1 --  39.1 \\
    & &        &           &                              & 39.1 -- 120.0 & \\
    & & no rot. & 8.0 -- 19.0 & 19.0 -- 32.0 & 32.0 -- 120.0  & \\
& \multicolumn{2}{l|}{This work (medium SN Ic\tablefootmark{2})} &  &  &  & \\
    & & rot. & 8.0 -- 16.8 & 16.8 -- 25.0 & 25.0 --  30.1 & 30.1 --  82.0 \\
    & &        &           &                   & 82.0 --  88.7 & 88.7 -- 120.0 \\
    & & no rot. & 8.0 -- 19.0 & 19.0 -- 32.0 & 32.0 --  52.2 & 52.2 -- 106.4 \\
    & &        &                   &                            & 106.4 -- 120.0 & \\
& \multicolumn{2}{l|}{This work (high SN Ic\tablefootmark{2})} &  &  &  & \\
    & & rot. & 8.0 -- 16.8 & 16.8 -- 25.0 & 25.0 --  29.0 & 29.0 -- 120.0 \\
    & & no rot. & 8.0 -- 19.0 & 19.0 -- 32.0 & 32.0 --  44.8 & 44.8 -- 120.0 \\
\cline{2-7}
\multirow{8}{0.5cm}{\rotatebox[origin=c]{90}{BH $\rightarrow$ no bright SN\tablefootmark{4}}$\left\{\rule{0cm}{1.45cm}\right.$}& \multicolumn{2}{l|}{\rule[0mm]{0mm}{3mm}This work} &  &  &  & \\
    & & no rot. & 8.0 -- 19.0 & 19.0 -- 32.0 & 32.0 -- 43.8 &\\
& \multicolumn{2}{l|}{This work (low SN Ic\tablefootmark{2})} &  &  &  & \\
    & & rot. & 8.0 -- 16.8 & 16.8 -- 25.0 & 25.0 -- 31.1 & 31.1 -- 33.9 \\
& \multicolumn{2}{l|}{This work (medium SN Ic\tablefootmark{2})} &  &  &  & \\
    & & rot. & 8.0 -- 16.8 & 16.8 -- 25.0 & 25.0 --  30.1 & 30.1 --  33.9 \\
& \multicolumn{2}{l|}{This work (high SN Ic\tablefootmark{2})} &  &  &  & \\
    & & rot. & 8.0 -- 16.8 & 16.8 -- 25.0 & 25.0 --  29.0 & 29.0 -- 33.9 \\
\cline{2-7}
\rule[0mm]{0mm}{3mm} & \multicolumn{2}{l|}{\citet{Georgy2009a}} &  &  &  & \\
    & & rot. & \multicolumn{2}{c}{8.0 -- 25.0} & 25.0 --  39.0 & 39.0 -- 120.0 \\
\hline
\end{tabular}
\tablefoot{All masses are given in $M_{\sun}$.\\
\tablefoottext{1}{Considering that the surface He abundance limit between WNL and WNE is $0.1$ instead of $10^{-5}$.}\\
\tablefoottext{2}{The maximum He mass ejected allowed to be still considered as a Type Ic SN is 0.4 - 0.6 - 0.8 $M_{\sun}$ in the low - medium - high SN Ic case respectively \citep[see][]{Georgy2009a}.}\\
\tablefoottext{3}{Assuming that the formation of a BH during the collapse has no influence on the SN explosion.}\\
\tablefoottext{4}{Assuming that the formation of a BH during the collapse prevents a bright SN explosion.}}
\end{table*}

\section{WR star populations: comparisons with the observations \label{SecPopu}}

\subsection{Positions in the HR diagram}

\begin{figure}
\centering
\includegraphics[width=.45\textwidth]{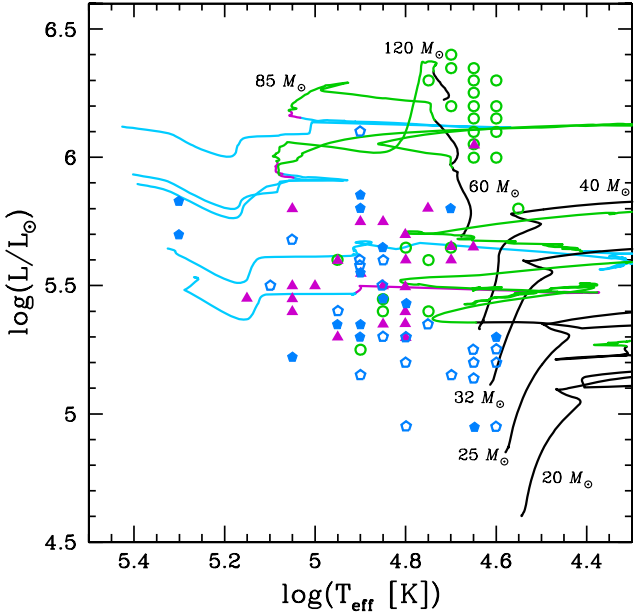}
  \caption{Positions of observed WN and WC stars in the HRD as given by \citet{Hamann2006a} and \citet{Sander2012a}, respectively. The empty circles are WNL stars, and the full triangles are WNE. The WC stars are represented by pentagons, filled when the distance is known and empty when it is unknown. The present rotating tracks are superposed.}
   \label{CompHR1}
\end{figure}

Observed positions of some galactic WN \citep{Hamann2006a} and WC  \citep{Sander2012a} stars are indicated in Fig.~\ref{CompHR1}. The luminosity of the most luminous WNL star in the sample of \citet{Hamann2006a} (not shown in Fig.~\ref{CompHR1}, see below) would require an initial mass of the order of 300 $M_{\sun}$. \citet{Crowther2010a} have recently claimed to have observed such massive objects. However, WR 25 (or HD 93162), the most luminous star, is a binary star with a spectroscopically determined orbital period of $207.8\, \text{d}$ \citep{Raassen2003a}. The second-most luminous star, WR 22, is an eclipsing binary. \citet{Rauw1996a} estimated a lower limit for the actual mass for this star to be 72 $M_{\sun}$. Owing to their binary status, these two stars are not plotted in Fig.~\ref{CompHR1}. Looking at Fig.~\ref{CompHR1}, a few points deserve to be emphasised:
\begin{itemize}
\item  The WNL, WNE, and WC present some kind of decreasing luminosity sequence. There are of course overlaps between the different WR types; however, just by eye one can notice immediately from Fig.~\ref{CompHR1} that the averaged luminosity of each class decreases from the WNL to the WC stars. The WR stars with the highest luminosity are therefore the WNL stars (the three upper points are also those with the highest abundances in hydrogen, with values between 0.35 and 0.53), and the star with the lowest luminosity is a WC star.
\item The range of effective temperatures (taken here at the surface of the hydrostatic core, \textit{i.e.} not accounting for the optical thickness of the wind) of the WNL stars  is relatively narrow (between 4.6 and 4.8) above a luminosity equal to about 5.8. Below, it widens and extends from 4.6 up to slightly more than 5. 
\item The ranges of effective temperatures covered by the WNE and WC are similar and extend towards higher temperatures than
the range of WNL stars. The difference in temperature ranges between the H-rich and the H-poor stars reflects  the strong dependency of the opacity on the quantity of hydrogen.
\end{itemize}

Before comparing with the present evolutionary tracks, let us try to explain the features listed above in a general theoretical framework. We focus here on the luminosity and exclude the question of the effective temperature. The reason for that is that the effective temperature depends a lot on the mass-loss rate used and on the physics of the outer layers, while luminosity is quite tightly related to the total mass of the star and its internal physics. In that respect, it is a more fundamental quantity to compare with models for the interior of stars.

The decrease in luminosity when passing from the WNL to the WC stars might be interpreted in two ways. To obtain a WC star from a given initial mass star, the star must lose more mass than what is required to obtain a WNL star, implying that on average WC stars have lower actual masses and consequently also lower luminosities. Another way  to interpret this feature (although not incompatible with the previous one) would be to assume that the most massive stars produce the high-luminosity WNL stars (which will then evolve into lower-luminosity WNE and WC-type stars), while the less massive stars produce the WC stars. In the framework of the single-star scenario, this last explanation requires very strong mass loss during the RSG phase of stars with initial masses of about $15\, M_{\sun}$. This strong mass loss could be caused by some physical processes originating in the envelope of the RSGs, while in the binary channel it could be caused by a RLOF process occurring in a close-binary system. Whatever process is invoked, it should not produce any long WNL or WNE phases at this low luminosity range since these stars are not observed. This point will be discussed in more detail in Sect.~\ref{SecSNe}.

\begin{figure}
\centering
\includegraphics[width=.45\textwidth]{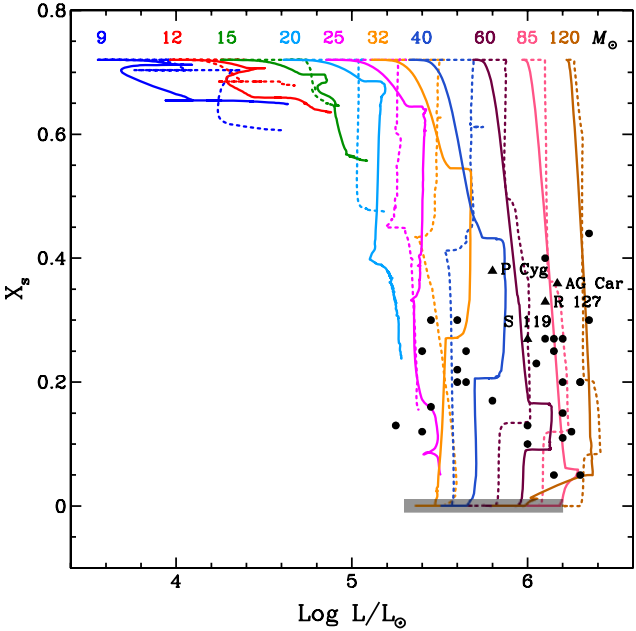} 
 \caption{Evolution of the mass fraction of hydrogen at the surface as a function of the luminosity. The continuous lines are the present rotating models, while the dotted lines represent the non-rotating ones. The dots are the WN stars with non-zero H-abundance and the shaded zone shows the range in luminosity of the WN stars with no H detected at the surface by \citet{Hamann2006a}. The triangles are LBV stars \citep{Groh2009b,Lamers2001a}. In this graph, models start at the top on the MS. They evolve to the right as their luminosity increases and downwards with mass loss peeling off the hydrogen-rich layers. The most massive models evolve back to the left (decreasing luminosity) due to the strong mass loss in the WR phase.}
 \label{Compxs1}
\end{figure}

\begin{figure}
\centering
\includegraphics[width=.45\textwidth]{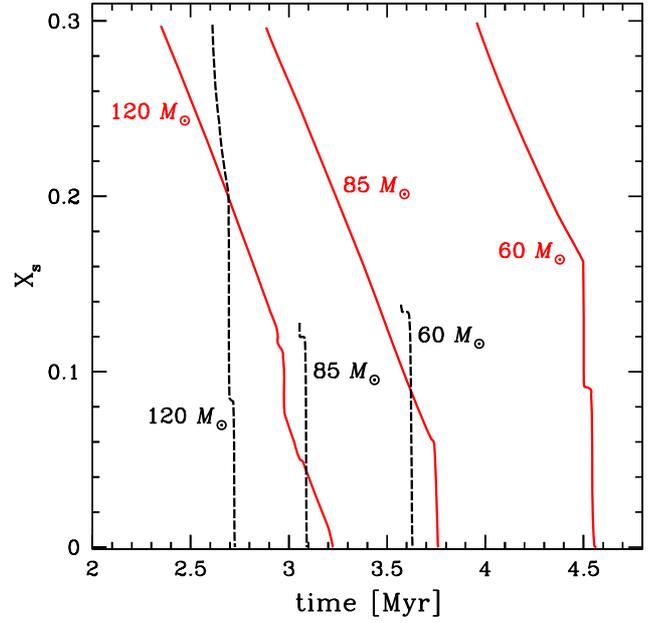} 
 \caption{Evolution of the mass fraction of hydrogen at the surface as a function of time during the WNL phase. The continuous red lines correspond to rotating models, while the dashed black lines correspond to non-rotating ones. The corresponding masses are indicated near the curves.}
\label{Compxs}
\end{figure}

Comparing the above observed positions with our rotating stellar models we note the following features. The group of seven very luminous stars ($\log (L/L_{\sun}) \geq 6.25$) all show a mass fraction of hydrogen higher than 10\%, except in one case where the mass fraction is estimated to be 0.05 (see Fig.~\ref{Compxs1}). In addition, no WNE and WC are observed in the luminosity range covered by these stars. Is there any explanation for these two features? The reason why only H-rich WN stars are observed in this high-luminosity range supports the idea that WR stars form through a combination of mass-loss and radiative-zone mixing, and not only through mass loss as in the non-rotating models. To illustrate this, we have plotted in Fig.~\ref{Compxs} the evolution as a function of time of the mass fraction of hydrogen at the surface of our $60$, $85$, and $120\, M_{\sun}$ stellar models with and without rotation during the WNL phase. We see that only the most massive star models with rotation ($M \geq 60\, M_{\sun}$) present an H-rich surface during sufficiently long periods (a few 100 thousands years) for allowing this phase to be observable. Corresponding models without rotation, or lower initial mass models, which enter into the WR phase only after the end of the MS phase, do not show any long ``H-rich'' periods and thus cannot account for the most luminous H-rich WNL stars. The absence of WNE and WC stars in this luminosity range may be explained by the fact, already mentioned above, that WNE and WC stars are expected to be less massive and therefore less luminous than WNL stars because more mass has to be removed from the star to reach those stages and/or because they are produced from stars with lower initial masses.

Figure~\ref{Compxs1} shows the evolution of the mass fraction of hydrogen at the surface as a function of the luminosity. Compared to non-rotating models, the rotating ones extend the regions covered by the WNL and WNE stars to a lower luminosity. However, the change remains modest so that comparisons with observations hardly allow one set to be favoured over the other. As emphasised above, the time spent in the H-rich portion of the diagram is probably much more decisive and favours the rotating models. We see that the tracks cover the region where the WNL stars are observed. The extension in luminosity of the WNE stars is also well reproduced, somewhat supporting the present single-star models for explaining these populations.

A serious difficulty is the low observed luminosity of some WC stars.  Our present tracks predict a lower luminosity limit for the WC stars of 5.35 (in $\log (L/L_{\sun})$), while the lowest luminosity plotted in Fig.~\ref{CompHR1} is about 4.9 according to the revised spectral analyses of galactic WC stars by \citet{Sander2012a}. We discuss this point in more detail below.

\subsection{Discussion of various possible origins for the low-luminosity WC stars}\label{SecDiscuWR}

What might be the origin of the observed low-luminosity (and thus low-mass) WC stars? If the luminosity is not underestimated at present, we can imagine three different kinds of evolutionary scenarios:
\begin{enumerate}
\item Stars with initial masses between $15$ and $20\,M_{\sun}$  lose much more mass than in the present grid of models during their RSG stage. We shall refer to this scenario as the \textit{RSG scenario}.
\item Stars with initial masses above about $25\, M_{\sun}$ lose much more mass than in the present grid of models (\textit{the massive star scenario}).
\item The low-mass WC stars are produced in close-binary systems through RLOF. This process can indeed produce lower final masses, as illustrated, for instance, by the recent models of \citet{Yoon2010a}. We shall refer to this scenario as the \textit{close-binary (CB) scenario}.
\end{enumerate}
Let us now discuss the advantages and disadvantages of each of these three scenarios.

\textit{1. The RSG scenario.} There are some mass-loss rate determinations for RSG pointing towards very high values. For instance, \citet{vanLoon2005a} obtained that dust-enshrouded RSGs present mass-loss rates that are a factor of 3--50 times higher than the rate of \citet{deJager1988a}. \citet{Humphreys1997a,Humphreys2005a} and \citet{Smith2009a} also indicated that the RSG star VY CMa went through important episodic mass ejections 500--1000 years ago. \citet{Moriya2011a} reached similar conclusions, studying the luminosity curve of SNe with RSG progenitors. Their results indicate that some mechanism is probably inducing extensive mass loss (greater than $10^{-4}\,M_{\sun}\,\text{yr}^{-1}$) in massive RSGs just before their explosions. \citet{Humphreys2008a} suggests that convective/magnetic activity may be the cause for what does appear as episodic and localised mass-loss events. Also very interesting are the observations by \citet{Davies2008a} of the so-called RSG clusters in the direction of Scutum. Maser emission indicative of a high-density medium above the photosphere is observed around the most luminous RSGs. This can be caused by the high mass-loss rates experienced by these stars. The presence of a luminous yellow supergiant in one of these clusters is also consistent with the idea that this star evolved away from the RSG stage \citep{Davies2008a}. 

\citet{Mauron2011a}, on the other hand, still recommend using the \citet{deJager1988a} rate for Galactic RSGs, indicating that this prescription agrees to within a factor of 4  with most mass-loss rate estimates based on the infrared 60 $\mu$m excess. This result is however compatible with the existence of short phases during which the mass-loss rates are much stronger. Indeed, mass-loss rates during the RSG phase may present some outbursting characteristics, somewhat similar (at lower luminosity and effective temperature) to what happens at higher luminosity and effective temperature for the luminous blue variable (LBV) stars \citep{Smith2011a}. The physical reasons for these high mass-loss rates may be related to the pulsational properties of RSGs \citep{Yoon2010b} and/or to the appearance of super-Eddington luminosities in the outer layers of the star \citepalias{Ekstrom2012a}.

From the theoretical point of view, it is well known that strong mass loss during the RSG phase favours a bluewards evolution \citep{Salasnich1999a,Vanbeveren1998b,Vanbeveren2007a} and thus helps in fulfilling one of the two minimal conditions for having a WR star, namely to have an effective temperature higher than about $\log (T_\text{eff}/\text{K}) = 4.0$. However, it is unclear whether this strong mass loss may lead to the formation of WC stars. \citet{Georgy2012a} has shown that increased mass-loss rates during the RSG phase (up to a factor of 10) are insufficient to lead to a WR star at the end of the evolution. To be valid, the RSG scenario should therefore involve a very strong increase of the mass-loss rate of more than 10 times during that phase with respect to the standard one.

One can already note that it will be difficult to separate the physical reason for the envelope loss, determining whether it is caused by a physical process occurring in the envelope of the star (as would be the case in the single-star channel) or to a RLOF. A constraint on this scenario may come from the fact that there are very few (if any) single-age clusters that simultaneously show RSGs and WR stars \citep{Humphreys1985a}, except in the clusters at the centre of our Galaxy \citep{Figer2007a} and in Westerlund 1 \citep[][although in the latter case we may see two clusters aligned instead of only one, as in the case of the Dansk clusters; see \citealt{Davies2012a}]{Negueruela2010a}. If some WR stars are produced from a range of initial masses that are also producing RSG, then to be consistent with this observation, either the RSG duration or the WR duration should be very short.

\textit{2. The massive star scenario.} In the high mass-loss rate grid of \citet{Meynet1994a}, a very low luminosity is reached for WC stars due to heavy mass loss. The lowest luminosity for the WC stars was at 4.5 and originated from the evolution of a $120\, M_{\sun}$! The high mass-loss rates used in \citet{Meynet1994a} are no longer supported by the more recent mass-loss determinations for O-type and WR stars that account for the effects of clumping \citep{Vink2001a,Nugis2000a}. However, very massive stars could lose very large amounts of mass in very short periods during which a strong mass outburst occurs \citep{Smith2006a}. Therefore, it is possible that the mass-loss determination, necessarily based on more frequent ``normal-moderate'' mass-loss rate stages, actually underestimates the true time-averaged mass-loss rate. 

\textit{3. The CB scenario.} In close binaries, the primary can lose its H-rich envelope during RLOF phases. The secondary can also undergo such a loss. The two stars may also merge or enter into a common-envelope phase leading to heavy mass loss episodes. Clearly, these evolutionary scenarios can lead to the production of WR stars and therefore have an impact on their populations. For instance, the close-binary scenario of \citet{Yoon2010a} makes it possible to produce final masses in the range between $1$ and $7\,M_{\sun}$, even when starting from high initial masses \citep[as for instance a $60\, M_{\sun}$, see][]{Yoon2010a}. However, according to the review by \citet[][see Fig. 4]{Crowther2007a}, the least massive WC star whose mass has been determined from binary orbit has a mass of about $9\, M_{\sun}$ (the most massive has about $16\, M_{\sun}$). The recent analysis of Galactic WC stars performed by \citet{Sander2012a} gives masses between $8$ and $37\, M_{\sun}$, covering quite well the masses obtained by our single-star models ($10$ -- $26\, M_{\sun}$). Therefore, the least massive He-core produced by the close-binary scenario cannot be invoked to explain the low-luminosity WC stars.

Close-binary evolution can produce WR stars from lower initial masses than the single-star evolutionary channel. For instance, \citet{Eldridge2008a} found that the minimum initial masses for stars to become WC stars is lowered from about $27\, M_{\sun}$ in the single-star scenario to $15\, M_{\sun}$ in the close-binary scenario. In the mass range between $15$ and $25\, M_{\sun}$, the WC stage occurs only at the very end of the evolution and has a duration of the order of $10^4$ years. This supports the view that low-luminosity WC stars could indeed be the result of close-binary evolution. It remains to be seen whether this scenario is able to explain the observed number of low-luminosity WC stars as well as their masses.

Accordingly at the moment, the three scenarios described above may contribute in explaining the low-luminosity WC stars. 
To investigate the relative weight of these different scenarios, progress must be made in several directions. We need to constrain the occurrence of short and strong mass-ejection episodes during the RSG phase. Could these events allow stars with initial masses around $15\, M_{\sun}$ to evolve into a WR phase?

Is there any evidence that the low-luminosity WC stars are produced in close-binary systems? Positive evidence could include the presence of a companion, or the determination that these stars are runaway stars, kicked off when the primary exploded in an SN event. Note however, that runaway stars may also be produced by dynamical interactions in dense clusters and thus result from processes other than binarity.

Is there any evidence for the presence of low-luminosity WC stars in very young associations, typically with a mass at the turnoff above about $25\, M_{\sun}$? These low-luminosity WC stars would originate from an initial mass more massive than $25\, M_{\sun}$ and this would support the second scenario (the massive star scenario).

\subsection{Surface chemical compositions}

\begin{figure*}
\centering
\includegraphics[width=.45\textwidth]{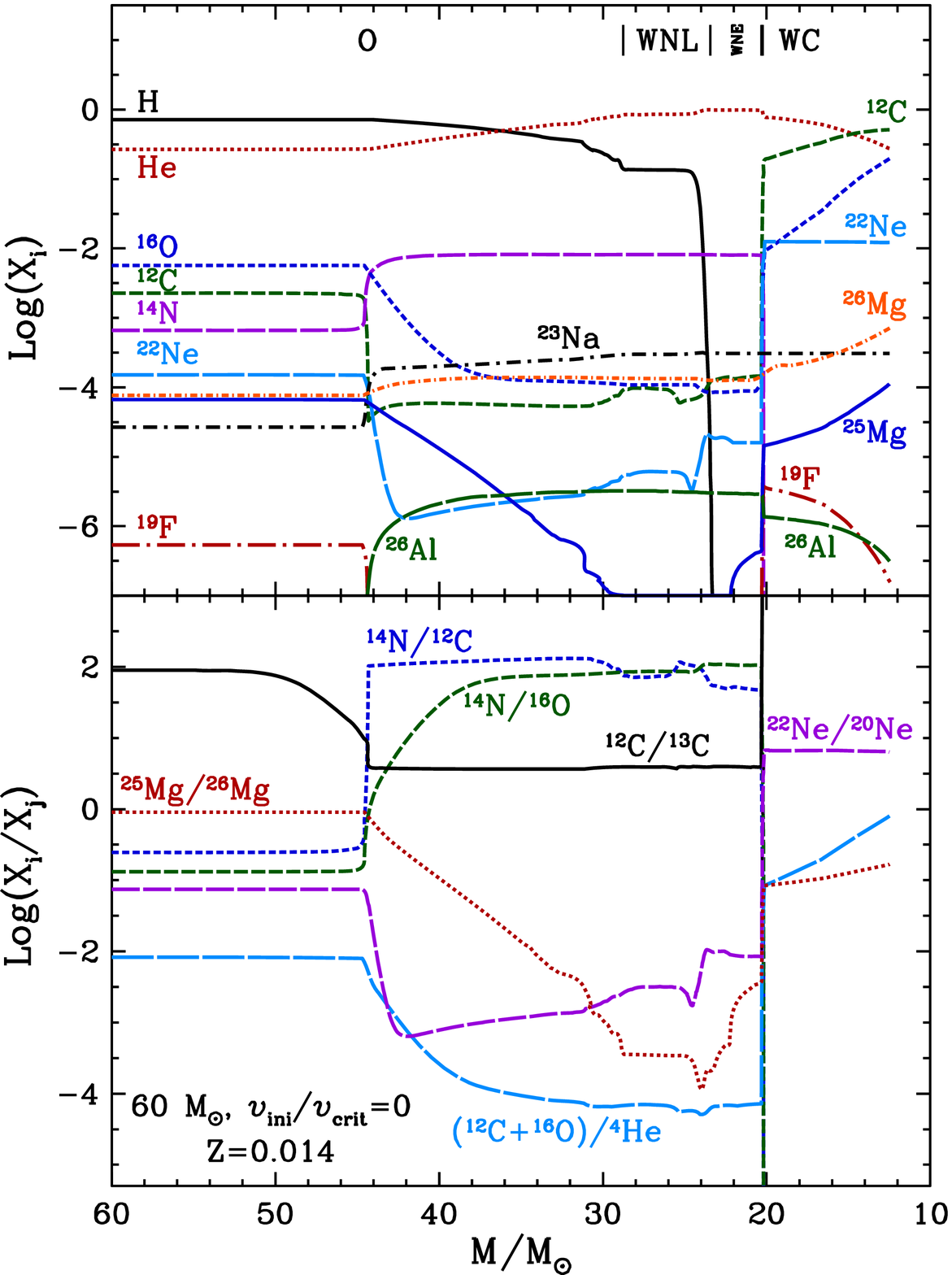}\hspace{.5cm}\includegraphics[width=.45\textwidth]{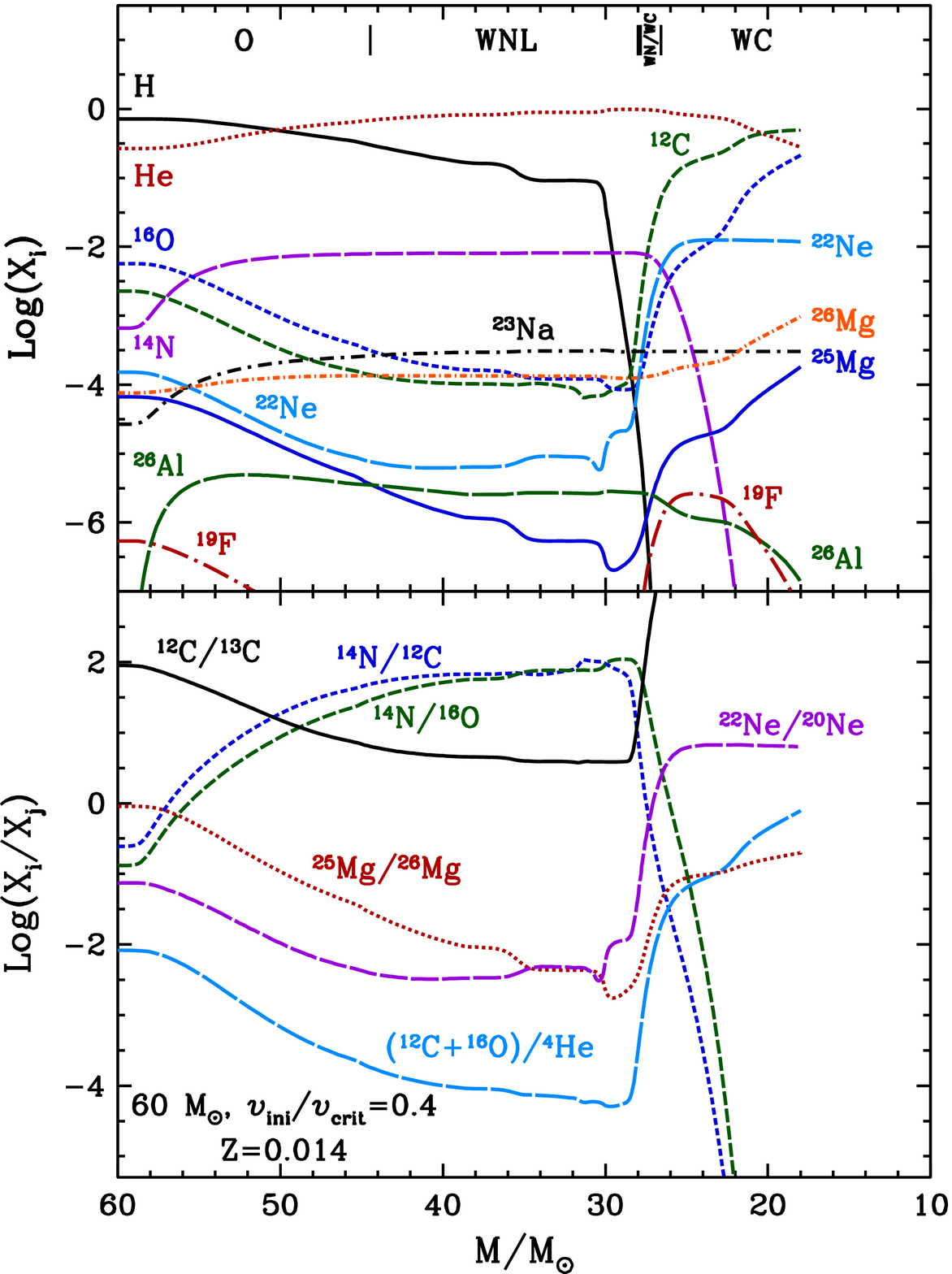}
  \caption{\textit{Top left panel:} Evolution of the surface abundances in mass fraction as a function of the actual mass of the star for the non-rotating $60\, M_{\sun}$ model. Different evolutionary phases are indicated in the upper part of the figure. A very narrow region, just before the WC phase, corresponds to the WNC phase (not labelled in the figure for clarity). \textit{Bottom left panel:} Evolution of abundance ratios (in number) as a function of the actual mass of the star. \textit{Right panels:} Same as the left panels for the rotating model.}
\label{c060S}
\end{figure*}

In Fig.~\ref{c060S}, the evolution of the surface abundances is shown for the  non-rotating and rotating $60\, M_{\sun}$ models. As already noted by \citet{Meynet2003a} \citep[see also][]{Maeder1987a,Fliegner1995a}, one of the main effects of rotation is to smooth the internal chemical gradients and to favour a more progressive arrival of internal nuclear products at the surface. This is well apparent comparing the curves shown in the left and right panels of Fig.~\ref{c060S}. 

One observable consequence is that rotating models predict more extended phases during which both H- and He-burning products are seen at the surface \citep{Langer1991a}. Note that in the left panel of Fig.~\ref{c060S}, the WNC phase is so narrow that it only appears as a thicker tick in the upper part of the top panel, while in the right panel of Fig.~\ref{c060S}, a well-extended phase is present. This phase overlaps with the WNE phase, which is not shown in the figure for clarity. 

Except for this effect, rotation leaves no other easily observable imprints on the way the surface composition evolves. This is quite expected because CNO equilibrium values are obtained during the WN phase that can be deduced from the nuclear properties of the chemical species and are not strongly affected by the details of the considered stellar model. 

This is also true for the abundance of $^{22}$Ne obtained during the WC phase. The abundance of $^{22}$Ne at this stage (WC) is actually an indication of the initial CNO content of the star. Indeed, $^{22}$Ne comes mainly from the destruction of $^{14}$N at the beginning of the helium-burning phase. This $^{14}$N is the result of the transformation of carbon and oxygen into nitrogen operated by the CNO cycle during the core H-burning phase. 

\begin{figure}
\centering
\includegraphics[width=.45\textwidth]{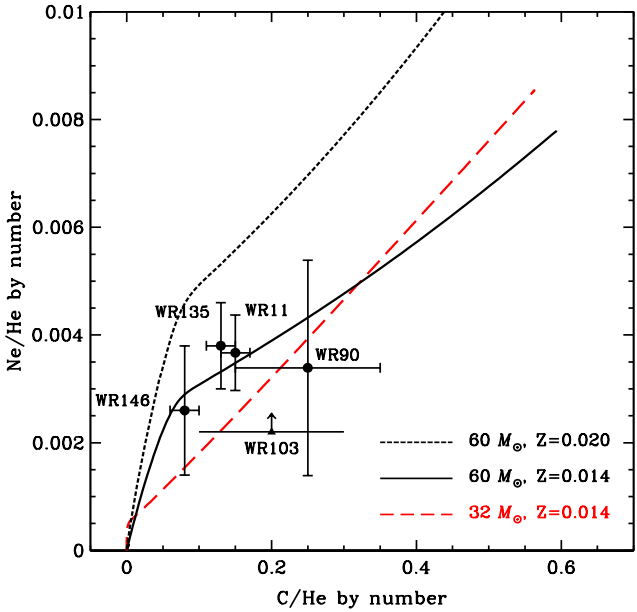} 
 \caption{Variations of the abundance ratios Ne/He vs. C/He at the surface of WC stars (in number). Models are represented by the lines (the Z=0.014 are the rotating models of the present paper, the $Z=0.020$ is the rotating model of \citealt{Meynet2003a}). The points are the observed values by \citet{Dessart2000a} (filled circles) and \citet{Crowther2006a}. }
 \label{neon}
\end{figure}

It is interesting to note that comparing the observed Ne/He ratios at the surface of WC stars with models computed with $Z=0.020$ shows that models over-predict the Ne abundance, while models starting with the solar abundances given by \citet{Asplund2005a} yield a much better fit, as can be seen in Fig.~\ref{neon}. This confirms that massive stars in the solar neighbourhood have initial metallicities that agree with the \citet{Asplund2005a} solar abundances. 

Let us note that this overabundance of $^{22}$Ne at the surface of WC stars is not only an important confirmation of the nuclear reaction chains occurring during He-burning, but is also related to the questions of the origin of the material accelerated into galactic cosmic rays  \citep{Binns2005a} and to that of the weak s-process in massive stars, since $^{22}$Ne is the main neutron  source in these stars \citep{Peters1968a}. 

Rotation increases the total quantity of ejected $^{26}$Al, as can be seen comparing the non-rotating and the rotating models. Note however, that the quantity of $^{26}$Al ejected by the winds of massive stars is lower in the present study than that found by \citet{Palacios2005a} for a given initial mass. This is a natural consequence of the lower ``solar" metallicity. Since the new solar abundances initially have fewer metals, this means that there is less $^{25}$Mg, the seed element for the synthesis of $^{26}$Al. The impact on the global production of $^{26}$Al in the Milky Way therefore needs to be revised. This will be done in a future paper. For now, we can say that the revised solar abundances diminish the role played by the WR stellar winds in the enrichment of the interstellar medium in $^{26}$Al.

As was proposed by \citet{Meynet2000b}, massive stars can also enrich the interstellar medium in fluorine through their winds. We confirm this with the present models \citep[see also][]{Palacios2005b}. Rotation does not appear to have a great impact on the wind ejection of this element.

\subsection{WR populations \label{SubSecWRPop}}

It is important to realise that to understand the origin of the WR populations, it is necessary to account for not only the relative number of WR of different subtypes with respect to the total number of WR, but also for the number of WR to O-type stars. For instance, if a good fit is obtained for the ratio WC/WR but the models only account for 10\% of the observed WR/O ratio, then this implies that the models predict a good WC/WR ratio in about 10\% of the WR stars with no guarantee that the fit will be good for the remaining 90\% of the WR stars. Therefore, it is very important to look at as many number ratios as possible to assess the quality of a model.

For the observed number ratio of WR to O-type stars, we focus on the spherical region of 2.5 kpc radius centred on the Sun for which extensive data are found in the literature. In the seventh catalogue of Galactic WR stars of \citet{vanderHucht2001a}, 64 WR stars were observed in that region. Ten years later, new detections by \citet{Pasquali2002a}, \citet{Negueruela2003a}, \citet{Hopewell2005a}, \citet{Hadfield2007a}, \citet{Shara2009a} and \citet{Mauerhan2011a} increased that number by about 20\% to 77. In that same zone the number of O-type stars is given to be equal to 481 in the \citet{Garmany1982a} catalogue. However, there is evidence that this number significantly underestimates the number of early-type stars in that region. For instance, more recent observations analysed by \citet{Reed2001a} indicate that the number of OB stars more massive than $20\, M_{\sun}$ within 2.5 kpc of the Sun might be 50\% greater than the value obtained by \citet{Garmany1982a}. In Table~\ref{TabWRSNrates} we use a number of O-type star within 2.5 kpc from the Sun increased by 33\% with respect to the number given by \citet{Garmany1982a} to account for some incompleteness. This gives a WR/O number ratio equal to 0.12 (=77/636). An error of about 25\% around this value is probably reasonable.

For the WR subtype populations, we consider the 3 kpc radius region centred on the Sun. Using the data collected in the above references, plus a few recent detections in the region between 2.5 and 3 kpc \citep{Mauerhan2009a,Roman-Lopes2011a,Anderson2011a}, there are 45 WN and 55 WC stars as different subtypes of WR stars. The 45 WN stars are distributed between 21 WNL, 23 WNE and 1 WNC stars. It must be emphasised here that the respective portion of WNL and WNE stars is quite uncertain. Where information on the mass fraction of H was available, we used the presence of H to classify a star as a WNL star and the non-detection of hydrogen for classifying it as a WNE star. For this we used the sample of Galactic WN stars analysed by \citet{Hamann2006a}, which has 20 stars in common with our sample of nearby WR stars. We classified in the remaining 25 stars all stars with an ``h'', indicative of the presence of hydrogen, as WNL, as well as all stars with spectral type equal or later than WN7; the remaining stars are considered to be WNE stars. Again, assuming an error of about 25\% we obtain the range of values given in Table~\ref{TabWRSNrates}. The WNC star is special in the sense that there is only one star of this type in the solar neighbourhood, which represents 1\% of WR stars in that same region. Among the WR stars presently detected in the Galaxy \citep[476 according to][but 548 if we add the 72 new WR detections by \citealt{Shara2011a}]{Mauerhan2011a}, 9 are at present identified as WNC stars, which corresponds to a proportion of about 2\%. Therefore we indicate a range between 1 and 2\% for this ratio in Table~\ref{TabWRSNrates}.

To compute the ratio of two types of stellar populations (A and B), we have to know the initial mass range leading to each population $\left[M_\text{min}^\text{A}\,...\,M_\text{max}^\text{A}\right]$ and $\left[M_\text{min}^\text{B}\,...\,M_\text{max}^\text{B}\right]$, as well as the time spent in the corresponding phase $\tau^\text{A}$ and $\tau^\text{B}$. When a constant star formation rate is assumed, the ratio of the two populations is given by
\begin{equation}
\frac{\text{N(A)}}{\text{N(B)}} = \frac{\int_{M_\text{min}^\text{A}}^{M_\text{max}^\text{A}} \phi(M)\tau^\text{A}(M)\text{d}M}{\int_{M_\text{min}^\text{B}}^{M_\text{max}^\text{B}} \phi(M)\tau^\text{B}(M)\text{d}M},\label{SNRatio}
\end{equation}
where $\phi(M)$ is the initial mass function. Here we consider a Salpeter initial mass function (IMF) \citep{Salpeter1955a}. The results are presented in Table~\ref{TabWRSNrates}.

The predicted value for the rotating models corresponds to only one initial $v_{\rm ini}/v_{\rm crit}$ value (0.4). A detailed computation should account for a distribution of the initial velocities. Since the value of 0.4 has been chosen to reproduce the average observed surface velocities during the MS for B stars \citep{Huang2010a}, we expect that the values of the ratios obtained for this particular value are close to the value that would be obtained by a properly weighted averaging over an initial velocity distribution (assuming that the characteristic initial ratio $v_\text{ini}/v_\text{crit}$ is the same for more massive stars). 

The value of the WR/O  ratio given by the non-rotating single-star model is 0.02, which is quite low compared to the observed ratio of around 0.12. The low ratio results from the low mass-loss rates used here that account for the clumping effect. If these models were the correct ones, then it would mean that close-binary evolution is responsible for more than 80\% of the WR stars. However, since rotating models better fit many observed features of massive stars (such as, for example, the change of the surface abundances), their predictions are to be preferred to those obtained from the non-rotating models. The WR/O value obtained by the rotating models, 0.07,  is more than three times the value obtained by the non-rotating models. It is below the observed values, which leaves some room for $\sim 40\%$ of binaries from the close-binary scenario to contribute to the number of WR stars at solar metallicity.

Since the WNL phase is increased in duration by rotation, the WNL/WR ratio is increased when the rotating models are used. The WNE/WR and WC/WR ratios, in contrast, decrease. The few WNC stars can be very well explained by internal mixing processes. 

To confront the present theoretical predictions with the observed numbers, we followed the same formalism as in \citet{Maeder1994a}. $\text{WR}_\text{s}$ is the number of WR stars that do not require any mass transfer episode for entering into their WR phase, and $\text{WR}_\text{cb}$ are those WR stars that owe their WR nature to a previous RLOF mass transfer. The total ratio WR/O can be written as the sum $\text{WR}_\text{s}/\text{O}+\text{WR}_\text{cb}/\text{O}$. $\varphi$ is the ratio $\text{WR}_\text{cb}/\text{O}$. Below we consider that the WNC  stars are produced by internal mixing only, not by a binary mass-transfer event.  In principle, a WNC star might result from a WN star that accretes matter from a WC star, but this does not appear to be realistic for at least two reasons: first, because of the compactness of these types of stars, some very peculiar initial conditions for the mass ratio and the period of the orbit are likely necessary; second, both types of stars suffer strong stellar winds, and processes such as wind collisions are more likely to occur than accretion. The WNC/WR ratio will therefore depend on $\varphi$ only through the dependence of the total number of WR. 

\begin{figure}
\centering
\includegraphics[width=.45\textwidth]{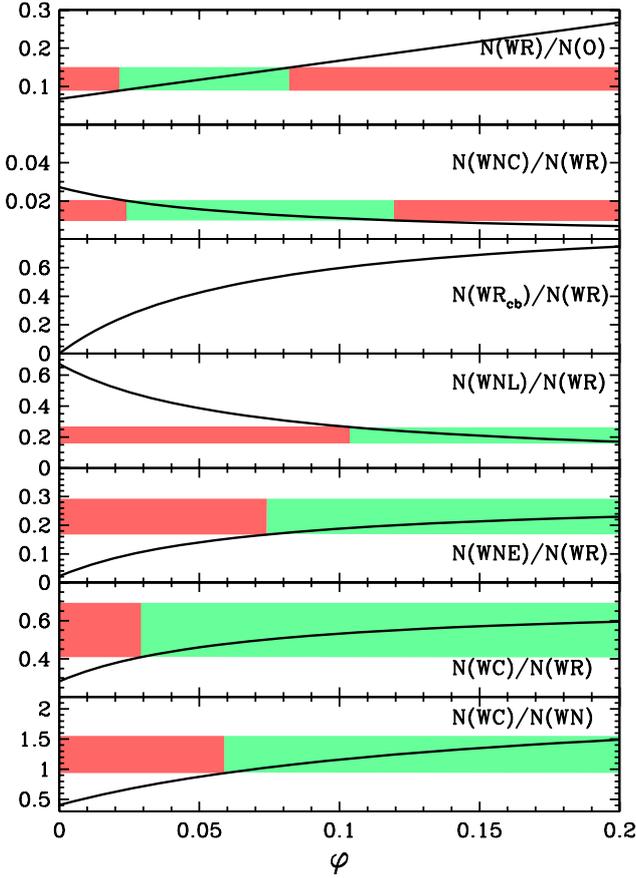}
   \caption{Variations of various number ratios as a function of $\varphi = \text{WR}_\text{cb}/\text{O}$, the fraction of WR stars with respect to O-type stars that owe their nature as WR stars to a RLOF episode in a close-binary system. The continuous lines in the 4 lower panels are the models for $\varphi_\text{WNL}$, $\varphi_\text{WNE}$, $\varphi_\text{WNC}$  and $\varphi_\text{WC}$ equal to $0$, $0$, $0.3\varphi$, and $0.7\varphi$ respectively (see text). The horizontal strips correspond to observed ratios as reported in Table~\ref{TabWRSNrates}. The green part corresponds to the values of $\varphi$ allowed by the observation, and the red one the excluded values.}
\label{comprap1}
\end{figure}

The first three upper panels of Fig.~\ref{comprap1} show how the WR/O, WNC/WR and $\text{WR}_\text{cb}/\text{WR}$\footnote{The fraction of WR due to close-binary evolution with respect to the total number of WR stars is given by $\varphi/(\text{WR}_\text{s}/\text{O}+\varphi)$.} compare to observations for different values of $\varphi$, using the results of the present rotating models for the fractions obtained when $\varphi =0$. We see that the range of values for $\varphi$ compatible with the observations is between 0.03 and 0.075, which corresponds to situations when 31 to 53\% of the WR stars owe their WR nature to a RLOF episode.

\begin{figure}
\centering
\includegraphics[width=.45\textwidth]{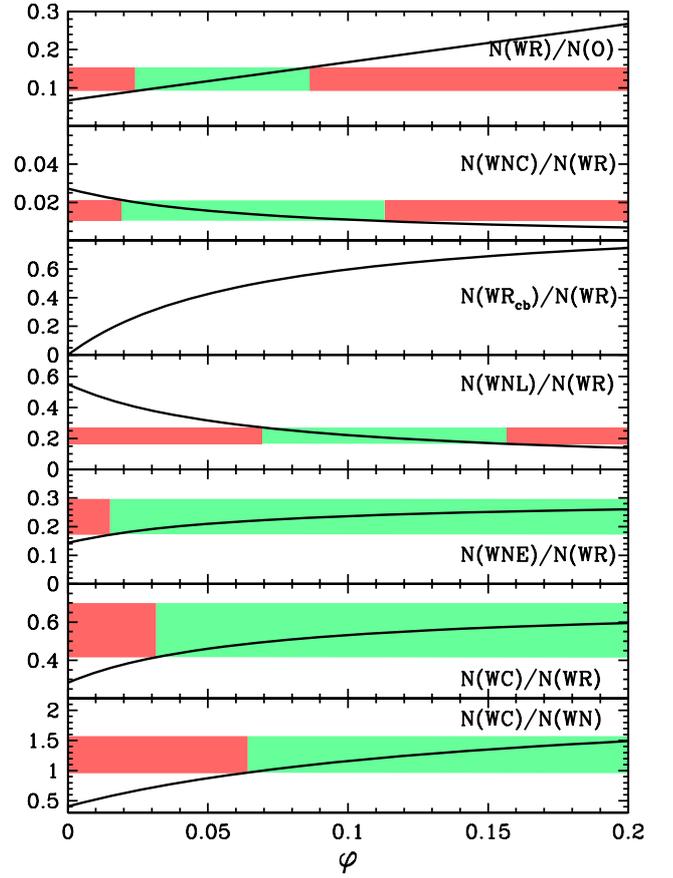}
   \caption{Same as in Fig.~\ref{comprap1}, but considering that WNE stars are WN stars with less than 10\% of hydrogen at the surface (mass fraction).}
\label{comprap2}
\end{figure}

To further pursue this line of reasoning and use comparisons with the observed WNL, WNE and WC star number ratios, we need to make some assumptions about the distributions in the WR subtypes resulting from the close-binary channel. We define $\varphi_\text{WNL}$ as $\text{WNL}_\text{cb}/\text{WR}_\text{cb}$. Analog definitions are considered for the other WR subtypes. With these definitions, the ratio of $\text{WR}(\text{st})/\text{WR}$ where ``st'' designates a given subtype can be written 
\begin{equation}
\frac{\text{WR}(\text{st})}{\text{WR}} = \frac{\frac{\text{WR}_\text{s}(\text{st})}{\text{WR}_\text{s}}+\frac{\varphi_\text{st} \varphi}{\text{WR}_\text{s}/\text{O}}}{1+\frac{\varphi}{\text{WR}_\text{s}/\text{O}}}.
\end{equation}
The values of the different components of  $\varphi$ should of course be the outcome of close-binary evolution calculations, such as those performed by \citet{Eldridge2008a}. Here, however, let us try a simpler approach guided by the comparisons of the observed ratios with the predictions of the rotating models.

For the reasons laid out above, we assume that $\varphi_\text{WNC}=0$. The WNL/WR ratio given by the present single-star rotating models is well above the observed range. This ratio therefore needs to be reduced when the close-binary scenarios are accounted for. To go in that direction we make the extreme assumption that $\varphi_\text{WNL}=0$. Even so, we see that a very high fraction of binaries would still be needed, which would be clearly incompatible with the observed WR/O.  This may indicate that the WNL phase is somewhat overestimated in the present models\footnote{We recall that here we only consider one value of $v_{\rm ini}/v_{\rm crit}$; the value obtained by a proper averaging over an initial velocity distribution would likely be lower than that obtained from the unique value of $v_{\rm ini}/v_{\rm crit}=0.4$}.

Considering that $\varphi_\text{WNE}=0.3\varphi$ and $\varphi_\text{WC}=0.7\varphi$, we find a consistent solution for $\varphi=0.075$ for all ratios shown in Fig.~\ref{comprap1}, with the exception of the WNL/WR ratio, which is above the observed range. At this point we conclude that mainly because of the high WNL/WR ratio obtained from the present models, we do not succeed in finding a completely consistent solution. To improve the situation, two lines of investigations can be followed, one observational and one theoretical.

From the observational point of view, it is difficult to match theoretical criteria and observational ones. For example, how much H can be hidden in an observed WNE? Obtaining new reliable measurements of the H abundance in all WN stars of the solar neighbourhood is crucial to find a better criterion for linking a given structure obtained in stellar models to an observed WR subtype. Some stars have changed classifications when more refined observations have been performed, indicating that the distinction between WNL and WNE is not always clear-cut. To illustrate how the WNL/WR and WNE/WR fractions depend on the definition of WNE, we indicate in Table~\ref{TabWRSNrates} the ratios obtained assuming that WNE stars are all WN stars with less than 10\% of hydrogen at the surface (``high-WNE case'', see Tables~\ref{TabMlim} and \ref{TabWRSNrates}). We also show in Fig.~\ref{comprap2} how the theoretical ratios compare with the observations. When adopting this new definition for the WNE stars, the WNE/WR ratio increases from 0.022 to 0.147, nearly a factor of 7. A consistent solution is now obtained for all ratios and for $\varphi \sim 0.07$, which corresponds to a share of about 50-50 between the single and binary channel for producing the observed WR populations at $Z=0.014$.

From the theoretical approach, one would explore a different mass-loss rate history during the WNL and WNE  phases. For instance, are the WN mass-loss rates the same when  the star is a core H-burning star or a core He-burning one? Are the mass-loss rates during the WNE phase overestimated at present? Stellar models in the future may explore the consequences of different prescriptions and offer some guidelines for improving the situations.

\begin{table*}
\caption{WR- and SN-type ratios deduced from our models, and comparison with previous works and observations.}
\label{TabWRSNrates}
\centering
\scalebox{0.9}{
\begin{tabular}{c r r| c c c c c c c}
\hline\hline
\rule[0mm]{0mm}{3mm}  & &  & WR/O-stars & WNL/WR & WNE/WR & WN/WR & WNC/WR & WC/WR & WC/WN \\
\cline{2-10}
\rule[0mm]{0mm}{3mm}  &  \multicolumn{2}{l|}{This work} &  &  &  &  &  &  & \\
    & & rot. & 0.066 &  0.687 &  0.022 &  0.709 & 0.028 &  0.291 & 0.409 \\
    & & no rot. &  0.015 &  0.253 &  0.168 & 0.421 & 0.003 & 0.579 & 1.376 \\
&  \multicolumn{2}{l|}{This work (High WNE)\tablefootmark{1}} &  &  &  &  &  &  & \\
    & & rot. & 0.065 & 0.562 & 0.147 & 0.709 & 0.028 & 0.291 & 0.410 \\
    & & no rot. & 0.015 & 0.185 & 0.238 & 0.423 & 0.003 & 0.577 & 1.364 \\
&  \multicolumn{2}{l|}{\citet{Meynet2003a}} &  &  &  &  &  &  & \\
    & & rot. & 0.07 & 0.66 & 0.05 & 0.71 & 0.04 & 0.25 & 0.35 \\
    & & no rot. & 0.02 & 0.35 & 0.16 & 0.51 & 0.00 & 0.49 & 0.97 \\
&  \multicolumn{2}{l|}{Observations} &  &  &  &  &  &  & \\
&  \multicolumn{2}{l|}{Solar Neighbourhood}& 0.12$\pm$0.03 & 0.21$\pm$0.05 & 0.23$\pm$0.06 & 0.44$\pm$0.11 & 0.015$\pm$0.05 & 0.55$\pm$0.14 & 1.25$\pm$0.30    \\
\cline{2-10}
\rule[0mm]{0mm}{3mm}     & &  & SN Ibc / SN II & SN Ib / SN II & SN Ic / SN II & SN Ic / SN Ib &  &   & \\
\cline{2-7}
\multirow{9}{0.5cm}{\rotatebox[origin=c]{90}{BH $\rightarrow$ bright SN\tablefootmark{3}}$\left\{\rule{0cm}{1.65cm}\right.$} &  \multicolumn{2}{l|}{\rule[0mm]{0mm}{3mm} This work (low SN Ic\tablefootmark{2})} &  &  &  &  &  &  & \\
    & & rot. & 0.241 & 0.186 & 0.054 & 0.290 &  &  & \\
    & & no rot. & 0.151 & 0.151 & -- & -- &  &  & \\
&  \multicolumn{2}{l|}{This work (medium SN Ic\tablefootmark{2})} &  &  &  &  &  &  & \\
    & & rot. & 0.241 & 0.066 & 0.174 & 2.635 &  &  & \\
    & & no rot. & 0.151 & 0.093 & 0.058 & 0.622 &  &  & \\
&  \multicolumn{2}{l|}{This work (high SN Ic\tablefootmark{2})} &  &  &  &  &  &  & \\
    & & rot. & 0.241 & 0.050 & 0.191 & 3.845 &  &  & \\
    & & no rot. & 0.151 & 0.066 & 0.085 & 1.279 &  &  & \\
\cline{2-10}
\multirow{8}{0.5cm}{\rotatebox[origin=c]{90}{BH $\rightarrow$ no bright SN\tablefootmark{4}}$\left\{\rule{0cm}{1.5cm}\right.$} & \multicolumn{2}{l|}{\rule[0mm]{0mm}{3mm}This work} &  &  &  &  &  &  & \\
    & & no rot. & 0.062 & 0.062 & -- & -- &  &  & \\
&  \multicolumn{2}{l|}{This work (low SN Ic\tablefootmark{2})} &  &  &  &  &  &  & \\
    & & rot. & 0.092 & 0.070 & 0.022 & 0.320 &  &  & \\
&  \multicolumn{2}{l|}{This work (medium SN Ic\tablefootmark{2})} &  &  &  &  &  &  & \\
    & & rot. & 0.092 & 0.061 & 0.031 & 0.521 &  &  & \\
&  \multicolumn{2}{l|}{This work (high SN Ic\tablefootmark{2})} &  &  &  &  &  &  & \\
    & & rot. & 0.092 & 0.050 & 0.042 & 0.857 &  &  & \\
\cline{2-10}
\rule[0mm]{0mm}{3mm}  &  \multicolumn{2}{l|}{\citet{Georgy2009a}} &  &  &  &  &  &  & \\
    & & rot. & 0.241 & 0.060 & 0.181 & 3.032 &  &  & \\
&  \multicolumn{2}{l|}{\citet{Meynet2003a}} &  &  &  &  &  &  & \\
    & & rot. & 0.32 &  &  &  &  &  & \\
    & & no rot. & 0.12 &  &  &  &  &  & \\
&  \multicolumn{2}{l|}{Observations} &  &  &  &  &  &  & \\
&  \multicolumn{2}{l|}{\citet{Boissier2009a}} & 0.14 -- 0.26 &  &  & 0.39 -- 1.13 &  &  & \\
&  \multicolumn{2}{l|}{\citet{Prieto2008a} ($Z\sim 1.3\, Z_{\sun}$)} & $0.572 \pm 0.286$ &  &  &  &  &  & \\
&  \multicolumn{2}{l|}{\citet{Prantzos2003a}} & $0.14$ -- $0.33$ &  &  &  &  &  & \\
\cline{2-10}
\end{tabular}}
\tablefoot{\tablefoottext{1}{Considering that the surface He abundance limit between WNL and WNE is $0.1$ instead of $10^{-5}$.}\\
\tablefoottext{2}{The maximum He mass ejected allowed to be still considered as a type Ic SN is 0.4 -- 0.6 - 0.8 $M_{\sun}$ in the low - medium - high SN Ic case, respectively \citep[see][]{Georgy2009a}.}\\
\tablefoottext{3}{Assuming that the formation of a BH during the collapse has no influence on the SN explosion.}\\
\tablefoottext{4}{Assuming that the formation of a BH during the collapse prevents a bright SN explosion.}\\
Remark: note that even though it seems that the non-rotating models better reproduce the various ratios of the WR subtypes, they are worse at reproducing the WR/O ratio. It is important to fit all these ratios \textit{simultaneously}.
}
\end{table*}

\section{Progenitors of type Ibc supernovae \label{SecSNe}}

The questions that we shall address in this section are the following ones:
\begin{itemize}
\item What are the predictions of the present models for the frequency of type Ib and Ic SNe?
\item How do these predictions compare to the observed ratios?
\item Can we find a consistent picture for understanding both the WR populations as observed in the solar neighbourhood and the frequency of type Ibc SNe, including contributions of close-binary evolution?
\item Which fraction of the present models would produce a fast rotating BH, accompanied by a type Ic SNe? The last question is triggered by the possibility that stars showing these conditions may give birth to a collapsar, proposed by \citet{Woosley1993a} as progenitors of LGRBs.
\end{itemize}

\subsection{Predictions of the models}

\begin{table*}
\caption{Properties of the models at the end of their evolution.}
\label{TabSNtype}
\centering
\begin{tabular}{r l| r r c r r r r r r c c}
\hline\hline
  \rule[0mm]{0mm}{3mm}   $M_\text{ini}$ & $v_\text{ini}/v_\text{crit}$ & $M_\text{He}$\tablefootmark{1} & $M_\text{CO}$ & $M_\text{rem}$ (baryon./grav.) & Tot. Ej. & H & He & Prog. type & SN type\tablefootmark{2} & Remnant & $P_\text{puls}$\\
  \rule[-1.5mm]{0mm}{3mm}  $M_{\sun}$ & & $M_{\sun}$ & $M_{\sun}$ & $M_{\sun}$ & $M_{\sun}$ & $M_{\sun}$ & $M_{\sun}$ & & & & $\text{s}$ \\
    \hline
  \rule[0mm]{0mm}{3mm}  120 & 0   & (30.91) & 30.13 & 9.13 / 4.51 & 21.72 & 0.00 & 0.67 &  WC & Ib & BH & -- \\
    & 0.4 & (19.04) & 18.46 & 5.69 / 3.65 & 13.29 & 0.00 & 0.52 &  WC & Ic & BH & -- \\
    85 & 0   & (18.65) & 17.98 & 5.54 / 3.60 & 13.09 & 0.00 & 0.47 &  WC & Ic & BH & -- \\
    & 0.4 & (26.39) & 25.72 & 7.88 / 4.26 & 18.47 & 0.00 & 0.61 &  WC & Ib & BH & -- \\
    60 & 0   & (12.50) & 12.23 & 3.91 / 2.91 &  8.57 & 0.00 & 0.42 &  WC & Ic & BH & -- \\
    & 0.4 & (17.98) & 17.52 & 5.40 / 3.55 & 12.58 & 0.00 & 0.50 &  WC & Ic+coll.? & BH & --\\
    40 & 0   & (12.82) & 10.08 & 3.40 / 2.64 &  9.41 & 0.00 & 0.95 & WNE & Ib & NS & -- \\
    & 0.4 & (12.33) & 12.21 & 3.90 / 2.90 &  8.42 & 0.00 & 0.41 &  WC & Ic+coll.? & BH & -- \\
    32 & 0   & (10.92) & 8.49 & 3.02 / 2.42 &  7.89 & 0.00 & 2.29 & WNE & Ib & NS & -- \\
    & 0.4 & (10.13) & 10.01 & 3.38 / 2.63 &  6.73 & 0.00 & 0.28 &  WC & Ic & NS & $1.3\cdot 10^{-4}$ \\
    25 & 0   & 8.12 & 5.95 & 2.41 / 2.03 &  5.85 & 0.03 & 2.20 & WNL & II-L/b & NS & -- \\
    & 0.4 & 9.69 & 7.09 & 2.69 / 2.22 &  6.99 & 0.00 & 1.59 &  WC & Ib & NS & $7.9\cdot 10^{-5}$ \\
    20 & 0   & 6.21 & 4.00 & 1.91 / 1.68 &  6.66 & 1.15 & 3.31 &  RSG & II-L/b & NS & -- \\
    & 0.4 & 7.17 & 4.73 & 2.10 / 1.82 &  5.06 & 0.02 & 1.61 & WNL & II-L/b & NS & $9.7\cdot 10^{-5}$ \\
    15 & 0   & 4.25 & 2.41 & 1.50 / 1.37 & 11.72 & 5.81 & 4.82 &  RSG & II-P & NS & -- \\
    & 0.4 & 5.11 & 3.19 & 1.71 / 1.53 &  9.36 & 3.31 & 3.75 &  RSG & II-P & NS & $9.2\cdot 10^{-5}$ \\
    12 & 0   & 2.99 & 1.75 & 1.33 / 1.23 &  9.97 & 5.64 & 3.75 &  RSG & II-P & NS & -- \\
    & 0.4 & 3.90 & 2.34 & 1.48 / 1.35 &  8.73 & 4.02 & 3.31 &  RSG & II-P & NS & $9.5\cdot 10^{-5}$ \\
    9 & 0   & 1.21 & 1.20 & 1.12 / 1.05 &  7.64 & 4.58 & 2.87 &  RSG & II-P & NS & -- \\
    & 0.4 & 3.08 & 1.64 & 1.30 / 1.20 &  7.22 & 3.53 & 3.03 &  RSG & II-P & NS & $1.2\cdot 10^{-4}$ \\
\hline
\end{tabular}
\tablefoot{The columns are: the initial mass (column 1), the initial velocity (column 2), the final masses of the He (column 3) and CO (column 4) cores, the baryonic and gravitational mass of the remnant (column 5), the total mass of the ejecta (column 6), the ejected mass of H (column 7) and He (column 8), the progenitor type (column 9), the SN type (column 10), the remnant type (column 11) and the pulsation period of the NS (column 12).\\
\tablefoottext{1}{The parenthesis indicates that the value corresponds to the whole stellar mass.}\\
\tablefoottext{2}{The SN types given here are determined using the ``medium SN Ic'' criterion ($M_\text{lim}^\text{He} = 0.6\, M_{\sun}$). The models that might lead to a LGRB through a collapsar event are indicated by the label ``coll.?'' (see text).}}\\
\end{table*}

To determine the type of the SN that occurs at the end of the evolution of our massive star models, we used the same procedure as in  \citet{Georgy2009a}. We recall here a few main points. First, we obtain the baryonic mass of the remnant using the same relation between the CO-core mass and the remnant baryonic mass as in \citet{Maeder1992a}. This mass is then used to compute the gravitational mass of the remnant, with the relation given in \citet{Hirschi2005a}. The results are given in Table~\ref{TabSNtype}. Using these data, we estimate the maximal mass on the zero-age main sequence (ZAMS) producing an NS during the SN event. As in \citet{Georgy2009a}, we consider the most massive NS to be $2.7\, M_{\sun}$. This choice is somewhat supported by the recent discovery of a very massive $2.4\, M_{\sun}$ NS by \citet{vanKerkwijk2011a}. We find that $M_\text{lim}^\text{NS-BH} = 33.9\, M_{\sun}$ (rotation), and $M_\text{lim}^\text{NS-BH} = 43.8\, M_{\sun}$ (no rotation). The difference is explained by the rotational mixing, which increases the size of the core. These limits are only poorly constrained, however, due to uncertainties during the explosion process, which are not accounted for in our simple estimate \citep[for example, the amount of matter that falls back, etc., see][]{Fryer2006a}.

Once the baryonic mass of the remnant is known, we deduce the composition of the ejecta, assuming that the entire mass between the surface and the edge of the remnant is ejected during the SN event. The results are shown in Table~\ref{TabSNtype}. In the same table, we also indicate the SN type, using the chemical composition of the ejecta and following the same criteria as in \citet{Georgy2009a}:
\begin{itemize}
\item If there is some hydrogen in the ejecta, the SN is a type II.
\item If there is no hydrogen and a helium mass higher than a given value ($M_\text{lim}^\text{He}$), it is a type Ib.
\item  If there is no hydrogen and a helium mass lower than $M_\text{lim}^\text{He}$, it is a type Ic.
\end{itemize}
Accounting for the uncertainty of $M_\text{lim}^\text{He}$, we give three values for the mass limit between types Ib and Ic SNe, with $M_\text{lim}^\text{He} = 0.4\, M_{\sun}$ (low SN Ic, as this criterion favours type Ib SNe), $M_\text{lim}^\text{He} = 0.6\, M_{\sun}$ (medium case), and $M_\text{lim}^\text{He} = 0.8\, M_{\sun}$ (high SN Ic, as it favours type Ic SNe).

Recently, \citet{Dessart2011a} found that some amount of He can be unobservable in the spectrum of an SN when it is located primarily in the most external layers (up to $50\%$ of He in mass fraction in the first $1\, M_{\sun}$ immediately below the surface). In this case, our non-rotating $120\,M_{\sun}$ and rotating $85\, M_{\sun}$ models would be classified as a type Ic SN instead of type Ib. This criterion thus produces the same mass limit between type Ib and type Ic SNe as in the ``high SN Ic'' case (see Table~\ref{TabWRSNrates}).

Contrarily to \citet{Georgy2009a}, where no distinction was considered between the various subtypes of type II SNe, we add in this work the same criterion as in \citet{Heger2003a} to distinguish between the type II-P SNe, and the type II-L or II-b SNe (owing to the weak statistics of these events, there is no simple way to distinguish between type II-L and II-b on the basis of the ejecta composition only. We therefore consider them as a unique sample):
\begin{itemize}
\item A type II SN is considered to be a type II-P if the ejecta contains more than $2\,M_{\sun}$ of H.
\item In the other case, it is considered to be a type II-L/b.
\end{itemize}
We see from Table~\ref{TabSNtype} that the upper mass limit for type II-P SN would be between $15$ and $20\, M_{\sun}$ for both non-rotating and rotating models. More precise mass limits can be obtained by interpolation. The mass limits become $19.0\, M_{\sun}$ for the non-rotating models and $16.8\, M_{\sun}$ for the rotating ones. Both values agree with the results by \citet{Smartt2009a}.

We computed the various SN type ratios with the same method as in \citet{Georgy2009a}, with the same IMF. We considered two cases: 1) a SN event is visible even when a BH is formed; 2) no visible SN occurs when a BH is formed. The results are shown in Table~\ref{TabWRSNrates}. They are quite similar to those obtained with different models by  \citet{Meynet2003a,Meynet2005a}, \citet{Georgy2009a}. Therefore, these results show some robustness against many changes in the physical ingredients of the models.

We see that rotation increases the SN Ibc/SN II ratio by about 60\% (48\%) in case 1 (2). The share of SN type Ibc between the Ib and the Ic types depends on the value chosen for $M_\text{lim}^\text{He}$. A low value implies more restrictive possibilities to obtain a SN Ic and consequently a lower SN Ic/SN II ratio and a higher SN Ib/SN II ratio. The trend of having more Ic SNe when rotation is accounted for remains true for all $M_\text{lim}^\text{He}$ considered here: rotation decreases the minimum initial mass of stars finishing their life as WC (see Table~\ref{TabMlim}). The effect of rotation on the SN Ib/SN II ratio is less clear: the ratio increases or decreases when rotation is accounted for depending on $M_\text{lim}^\text{He}$. 

\subsection{Comparisons with observations}

At solar metallicity, in case 1, the present models give ratios well within the range of the observed values determined by \citet{Boissier2009a}. Indeed, one sees that the present single-star models may explain the greatest part of the observed number of type Ibc  (normalised to the number of type II SNe). For the SN Ic/SN Ib ratio, as already noted above, the value greatly depends on the value of $M_\text{lim}^\text{He}$.  We can only say that to achieve a reasonable fit, rotating models would favour a low-medium value while the non-rotating ones favour a medium-high value.

In our case 2, about half of the type Ibc SNe occurring at solar metallicity could originate from the single-star channel \citep[similar to the conclusion we reached in][]{Georgy2009a}. The majority of the type Ibc SNe produced by the single-star channel would be type Ib SNe. If the hypothesis ``formation of BH implies no visible SN'' is correct, a significant fraction of the Ibc SNe must be produced in scenarios different from those computed in the present grid of models. They must originate from stars producing sufficiently low-mass cores to avoid BH formation. Note also that if the ``massive star'' scenario is correct (see Sect.~\ref{SecDiscuWR}), this would lead to less massive cores, and thus favour the production of NS instead of BH.

The idea of very low-luminosity progenitors for type Ibc SNe has recently received some support by \citet{Smartt2009b}. This review highlights that no progenitors have been detected at the positions of ten type Ibc SNe for which images have been obtained prior to the SN event. As indicated by this study, such a non-detection cannot rule out a massive WR star progenitor. However, under the hypothesis that the progenitor population of all Ibc SNe are massive WR stars, the probability that they did not detect any of the 10 progenitors by chance would then be only 11\%, assuming that the Ib progenitors are WN stars and Ic progenitors are WC/WO stars \citep[see more detail in][]{Smartt2009b}. Of course, uncertainties about the distance and the estimated degree of absorption (owing, for instance, to an enhanced mass loss episode just before explosion) may play a role when rejecting typical WR stars as good candidates for the progenitors of type Ibc SNe.

Let us suppose, for discussion purposes, that at least some of the progenitors of type Ibc SNe could be low-mass, low-luminosity, naked CO-core stars. We may first wonder to which extent such a population is observed. Could it correspond to the low-luminosity WC stars discussed above? The answer is likely no, since many of the low-luminosity WC stars analysed by \citet{Sander2012a} have visual magnitudes well above the lower visual magnitude, which would imply detection of the progenitor by  \citet{Smartt2009b}. This means that the progenitors of these type Ibc could be completely different from the WR stars we observe, even at the lowest luminosity.

Why do we not observe such progenitors? Where are they? They may have escaped detection in the case of the ten Ibc SNe because they are too distant in these particular cases, but how can we explain the fact that this population has escaped detections independent of any link with particular SN events? He- or CO- rich stars of a few solar masses are luminous enough to be detected as individual stars. If they are not observed, an explanation has to be provided. Are they hidden in the light of their companion in a close-binary system? Do they escape detection because they may be enshrouded in heavy circumstellar material coming from their companion, or from matter lost by the system? Is their lifetime as a naked He- or CO-core so short that they are very seldom caught in that stage of their evolution? These questions remain quite open at the moment.

If these stars exist, they would likely be produced in close-binary systems. As mentioned above, the RLOF process makes it possible to produce very low final masses between 1 and $7\,M_{\sun}$, even when starting from high initial masses \citep[for instance a $60\, M_{\sun}$, see][]{Yoon2010a}. It also makes it possible to produce naked He-cores from lower initial-mass stars. Close-binary scenarios of \citet{Yoon2010a} actually predict a bimodal distribution for the progenitors of type Ic: the high-mass progenitors (initial masses greater than  $35$ -- $45\, M_{\sun}$) and the low-mass progenitors (between $12.5$ -- $13.5\, M_{\sun}$). However, this conclusion is quite dependent on the assumption made for the maximum quantity of helium in the ejecta, that is still compatible with a type Ic event (above numbers are obtained for  a maximum value of $0.5\, M_{\sun}$).

Can we conclude that single-star models would produce most of the observed WR stars, while close binaries would produce the majority of type Ibc? We think that this is likely too schematic a representation of reality. First, not all single-star models will produce a BH (even with the weak mass-loss rates used in the present work) and therefore some single stars will produce a visible type Ibc event. Second, in the absence of any reliable understanding of how massive stars explode (particularly if rotation is involved), it may be premature to draw firm conclusions regarding the nature of the visible counterpart of the final core-collapse. Third, the uncertainties concerning the mass-loss rates have a strong impact on both single and binary scenarios and still prevent definitive conclusions from both channels. Fourth, it would be striking that the predicted ratios for type Ibc SNe from single-star models would fall so well in the range of the observed values just by chance. Finally, studying the distribution of WR stars with respect to their host light distribution, \citet{Leloudas2010a} found similar trends between the distributions of WN and WC stars compared to those of SNe Ib and Ic, supporting the idea that WR stars are the progenitors of at least part of the type Ibc SNe. For all these reasons, we think that any strong statement attributing all Ibc events to close-binary evolution or to single stars is quite premature at present. Nature probably includes both channels, but with a frequency which still remains difficult to assess quantitatively.

\begin{figure}
\centering
\includegraphics[width=.45\textwidth]{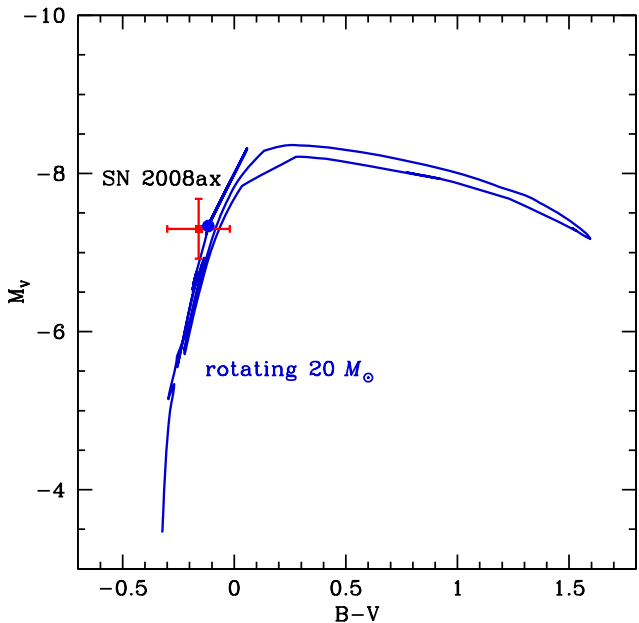}
\caption{Track of the rotating $20\, M_{\sun}$ model in the $B-V$ vs $M_V$ plane. The end point of the track is the blue point. The red square corresponds to the estimated position of the progenitor star of SN 2008ax \citep{Crockett2008a}.}
\label{ColMag20}
\end{figure}

In this context, it is very interesting to mention two recent observations regarding progenitors of core collapse SNe: 1)  \citet{Smartt2009a} found no RSG type II-P progenitors in the mass range between about 18 and $25\, M_{\sun}$. In our opinion, this may well be explained by a strong mass loss during the RSG stage, which would lead stars in this mass range to explode as a type II-L, type II-b, type Ib, or type Ic SN (see Table~\ref{TabSNtype}). Of course, the strong mass loss could be stimulated by the presence of a companion, but in that case one should provide an explanation for why it only occurs above a certain mass limit. At the moment, we tend to favour an explanation based on some physical process becoming active only above a mass-luminosity threshold, and related to pulsation \citep{Yoon2010b} or supra-Eddington luminosity \citepalias{Ekstrom2012a}. 2) Observations of yellow supergiant progenitors for IIb SNe have been made recently \citep[][see \citealt{Maund2011a} and references therein]{Crockett2008a}. These progenitors likely evolved back from a RSG stage, and might be explained by some enhanced mass loss during the RSG stage. This supports the statement above and is well in line with the maser observations by \citet{Davies2008a} discussed in Sect.~\ref{SecPopu}. This possibility has recently been discussed by \citet{Georgy2012a}.

Let us mention here the particularly interesting case of SN 2008ax discussed in \citet{Crockett2008a}. The authors mention the possibility that the progenitor of this SN could be a WNL star, with a very low hydrogen surface abundance. However, their non-rotating models do not reproduce the position of the observed progenitor star in the HRD well. Their best estimate is a $\sim 27$ -- $28\, M_{\sun}$ model. The difficulty of this model is the incongruously high mass of their CO core \citep[around $9\, M_{\sun}$, while the light curve and ejecta velocities of the SN are compatible with a core mass of about $4$ -- $5\, M_{\sun}$; see][]{Crockett2008a}. As shown in  Fig.~\ref{ColMag20}, the track of our rotating $20\, M_{\sun}$ model in the $B-V$ vs $M_V$ plane\footnote{\footnotesize{The conversion of the quantities $\log(L)$ and $\log(T_\text{eff}/\text{K})$ to $M_V$ and $B-V$ is done according to \citet{Boehm-Vitense1981a}, \citet{Flower1977a}, \citet{Sekiguchi2000a}, \citet{Malagnini1986a} and \citet{Schmidt-Kaler1982a}.}} ends noticeably close to the SN progenitor position. Moreover, it ends as a WNL star (see Table~\ref{TabWRtau}), with a H-content in the ejecta of $0.02\, M_{\sun}$, and a CO-core mass of $4.73\, M_{\sun}$ (see Table~\ref{TabSNtype}). It is thus an excellent candidate for explaining the observed properties of the progenitor of SN 2008ax!

\begin{figure*}
\centering
\includegraphics[width=.45\textwidth]{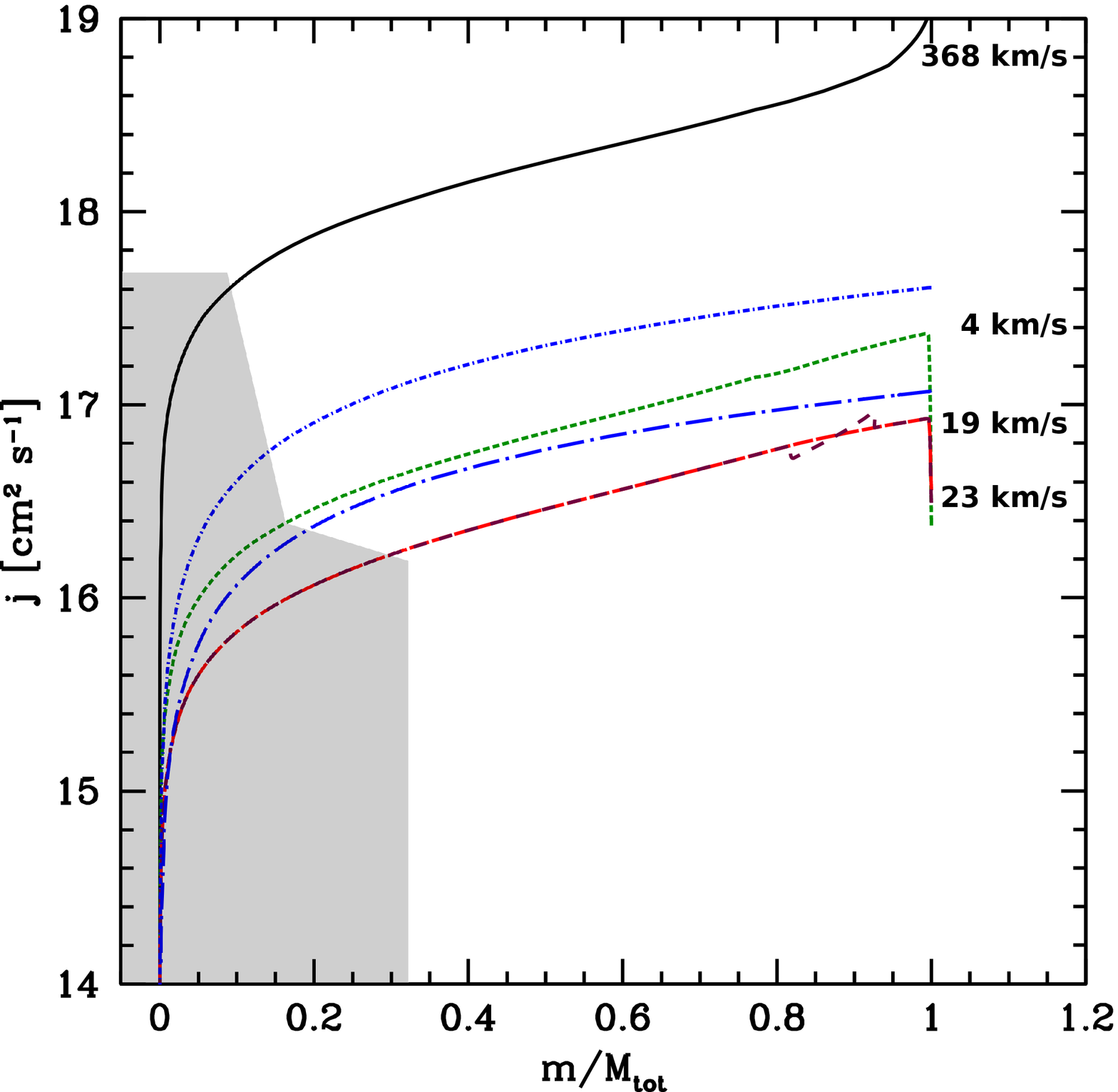}\hspace{.5cm}\includegraphics[width=.45\textwidth]{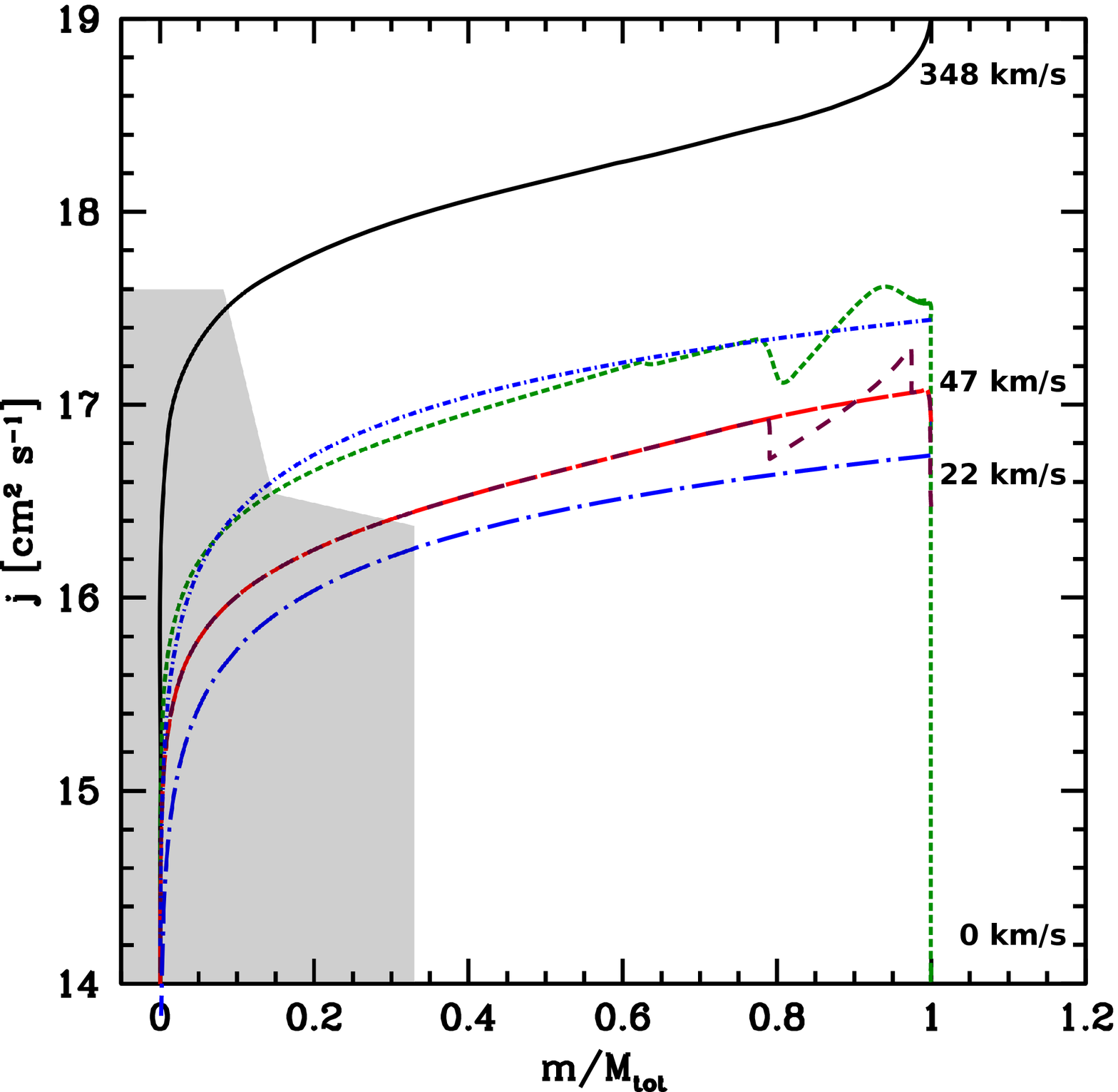}
\caption{Specific angular momentum for the rotating $85\, M_{\sun}$ (\textit{Left}) and $60\, M_{\sun}$ (\textit{Right}) models as a function of the Lagrangian mass coordinate normalised to the actual total mass of the star. The continuous (black) curve corresponds to the ZAMS, the short-dashed (green) curve to the beginning of the core He-burning phase, the long-dashed (red) curve to the beginning of the C-burning phase, and the medium-dashed (purple) curve to the end of the C-burning phase. The dot-short-dashed (blue) curve indicates the minimum specific angular momentum for a stable circular orbit around a Schwarzschild black-hole and the dot-long-dashed (blue) indicates the minimum specific angular momentum for a stable circular orbit around a maximally rotating Kerr black-hole. The grey area cuts the lines at the point corresponding to the mass coordinate equal to the mass of the remnant indicated in Table~\ref{TabSNtype}. The surface equatorial velocities are indicated for each stage.}
\label{FigMomCin}
\end{figure*}

\section{Rotation rate of young pulsars and progenitors of LGRB}\label{SecGRB}

\begin{figure*}
\centering
\includegraphics[width=.45\textwidth]{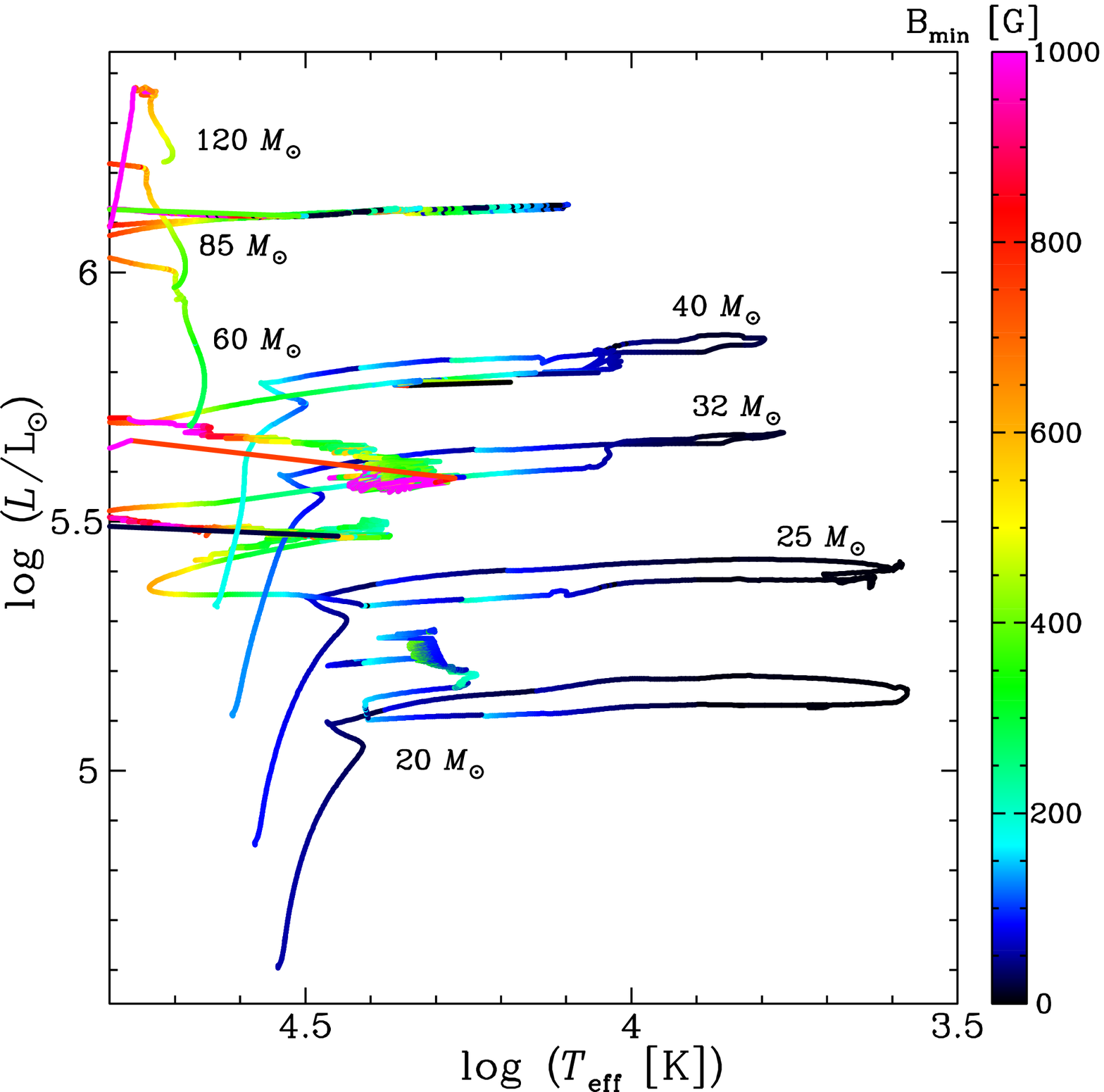}\hspace{.5cm}\includegraphics[width=.45\textwidth]{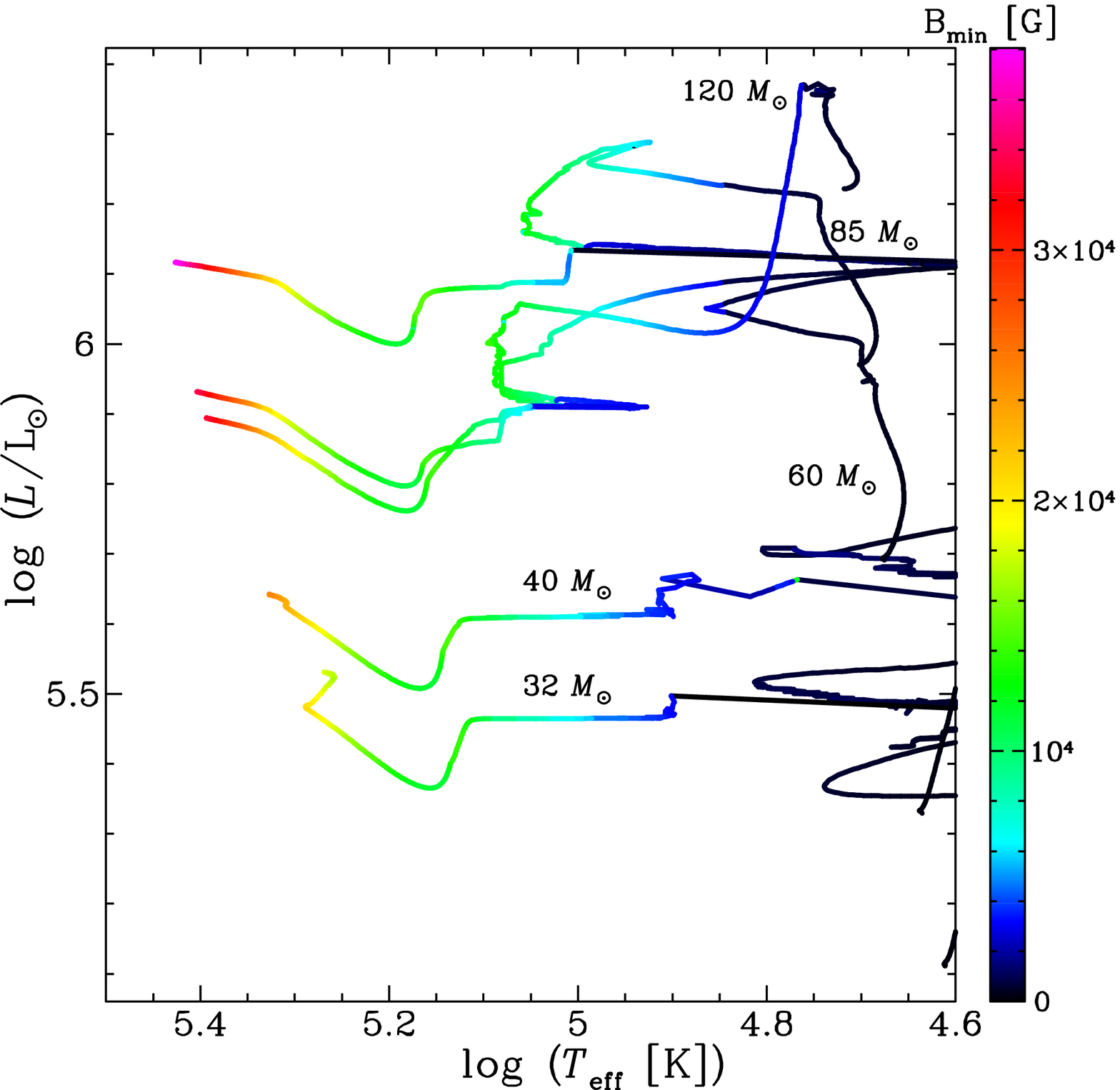}
\caption{Evolutionary tracks in the HRD. The colours along the tracks indicate the minimum magnetic field (assumed to be aligned with the rotation axis and dipolar) able to couple the stellar wind with the stellar surface and thus to exert a torque. Higher magnetic fields than the minimum value indicated here would reduce the angular momentum content of the star. The minimum value is obtained by imposing that the parameter $\eta$ defined in \citet{ud-Doula2009a} is equal to one. {Left panel:} Main-Sequence and RSG phases. {Right panel:} WR phase.}
\label{BminRSG}
\end{figure*}

Figure~\ref{FigMomCin} shows how the specific angular momentum varies inside the rotating models at different evolutionary stages. It should be kept in mind that in the absence of any transport of angular momentum, the specific angular momentum would remain constant with time. In the current models, of course, transport mechanisms triggered by convection and rotation are present and therefore change the specific angular momentum. As a numerical example, we consider our $60\, M_{\sun}$ model (right panel). At the beginning of the core H-burning phase, the specific angular momentum at the mass coordinate corresponding to the remnant mass is $3.2\cdot 10^{17}\,\text{cm}^2\,\text{s}^{-1}$. At the end of the core H-burning phase, the specific angular momentum at the same point is $3.6\cdot 10^{16}\,\text{cm}^2\,\text{s}^{-1}$, \textit{i.e.} decreased by nearly a factor of 9. At the end of our computation, just after the core C-burning stage in this case, the specific angular momentum is $2.5\cdot 10^{16}\,\text{cm}^2\,\text{s}^{-1}$, \textit{i.e.} decreased by nearly a factor of 13 with respect to the initial value. Since the evolution is very rapid up until the SN explosion, this number is likely to be a very good approximation of the specific angular momentum at the time of the pre-SN stage and at the border of the region that will be locked into the stellar remnant. Interestingly, it appears that more than 96\% of the reduction of the specific angular momentum at that mass coordinate occurs during the MS phase. This underlines, as already emphasised in previous works \citep[see \textit{e.g.}][]{Hirschi2004a,Hirschi2005a,Woosley2006b}, how important it is to properly account for angular momentum transfer during the MS phase. 

Another point worth noting is that a low surface velocity does not necessarily imply a low angular momentum for the central region of the star. Typically, at the end of the evolution of our $60\, M_{\sun}$ model (see Fig.~\ref{FigMomCin}), the surface equatorial velocity is only $22\,\text{km}\, \text{s}^{-1}$, while the specific angular momentum at the edge of what will become the stellar remnant is well above the minimum required for a stable orbit around a Kerr BH, {\it i.e.} sufficient for allowing the accreting matter to form an accretion disk around the BH that will likely form at the end of the stellar lifetime.

Assuming that the angular momentum content in the mass coordinate that will become the remnant remains in the NS, we then obtain the pulsar period at birth indicated in Table~\ref{TabSNtype}\footnote{To compute the periods, we assumed that all the angular momentum contained in the Lagrangian mass $M_\text{rem, bar}$ remains in the NS. We then used the relation $\mathcal{I} = 0.35\, M_\text{rem, grav}R^2$ where $\mathcal{I}$ is the momentum of inertia of the NS \citep{Lattimer2001a}, and where a radius $R=12\,\text{km}$ is assumed for the NS.}. The periods for our models range between 0.08 and 0.13 ms. These are extremely short periods, not only much shorter than the observed periods of young pulsars, which are between 20 and 100 ms \citep{Muslimov1996a,Marshall1998a,Kaspi1994a}, but even shorter than the critical velocity of NS, which is between 0.44 and 0.65 ms\footnote{We estimated this critical period on the basis of the gravitational mass of our remnant, a radius of $12\,\text{km}$ for the NS, and a critical (Keplerian) angular velocity given (in a classical way) by $\Omega_\text{crit} = \sqrt{\frac{GM}{R^3}}$.} Some additional braking mechanism is therefore needed to reconcile our predictions with observations (see below).

A similar conclusion can be obtained by considering the frequency of LGRBs. Assuming that a LGRB is produced every time the three following conditions are fulfilled: 1) a BH is formed, 2) the angular momentum of the newborn BH is high enough to form an accretion disk, 3) a type Ic SN is associated with the event. These conditions are those needed to obtain a collapsar \citep{Woosley1993a}, and to the observational association of broad-lined Type Ic SNe with several nearby ($z \le 0.5$) LGRB events \citep[][and references therein]{Woosley2006a,Berger2011a}. With this hypothesis, we obtain favourable conditions for a collapsar in the mass range between 40 and $60\, M_{\sun}$. In Table~\ref{TabGRB} we show the expected ratio of collapsar to core-collapse SN (CCSN) obtained from present rotating models, computed using Eq.~(\ref{SNRatio}). They are between 5 and 9\% of the rate of core-collapse SNe. The fraction of LGRB to SN Ibc is much higher than the recent estimate of this fraction by \citet{Soderberg2010a}, which is about 1\%, and is obtained from a radio survey of type Ibc SNe\footnote{This technique allows the detection of undetected LGRBs, since any relativistic outflow will produce a long-lived non-collimated afterglow.}.

\begin{table*}
\caption{Mass range and ratio of LGRB compared to various SN types.}
\label{TabGRB}
\begin{center}
\begin{tabular}{c|cc|cc}
\hline\hline
mass range & \multicolumn{2}{c|}{BH $\rightarrow$ bright SN} & \multicolumn{2}{c}{BH $\rightarrow$ no bright SN}\\
 & LGRB / CCSN & LGRB / SN Ibc & LGRB / CCSN & LGRB / SN Ibc \\
 \hline
 \rule[0mm]{0mm}{3mm}$40.0$ -- $60.0\, M_{\sun}$ & 0.049& 0.254 & 0.056 & 0.663 \\
  \rule[-1.5mm]{0mm}{3mm}$36.0$ -- $67.0\, M_{\sun}$ & 0.076 & 0.394 & 0.087 & 1.029 \\
\hline
\end{tabular}
\end{center}
\end{table*}

From the above discussion, it appears that some additional braking mechanism is at work. What could this additional braking mechanism be, and when does it intervene during the evolution of the star? Such a braking mechanism could intervene during the nuclear lifetime of the star, at the time of the core-collapse event and/or during the very first phases after the birth of the neutron star.

At the moment, the solution that has been extensively explored in the literature invokes a strong coupling mechanism active inside the star during its nuclear lifetime. This coupling could be caused by a strong interior magnetic field imposing a solid body rotation law during the core H-burning phase. The interior magnetic field could be fossil or generated by a shear-dynamo in the radiative layers. Coupled with the mass loss prescriptions used in stellar evolution, these models can reproduce the shortest observed periods of young pulsars \citep{Heger2005a} and restrict the progenitors of LGRB to those stars beginning their evolution with a very high initial velocity, so high that they follow a nearly chemical homogeneous evolution  \citep[see e.g.][] {Woosley2006b,Yoon2006a}. Moreover, these models obtain favourable conditions for LGRB only at low metallicities.

This solution is quite elegant in the sense that the physics of stars becoming pulsars (or NS) and stars becoming LGRB is the same (all have nearly solid body rotation during the MS phase). Only the initial velocities are different. However, there are also problems with this solution. First, there are still difficulties in accounting for young pulsars with the longest periods, indicating that even in that case an additional braking mechanism may be missing. Second, observations from asteroseismology indicate that a strong gradient of the angular velocity seems to be present in some massive stars (even stronger than in the present models), showing that at least some massive MS stars avoid  solid body rotation \citep{Aerts2008a}. Third, some authors \citep{Zahn2007a} have challenged the shear dynamo theory presently adopted in the models responsible for the strong coupling. Fourth, if the field is fossil, one can wonder why it has no surface components. A surface magnetic field may in the latter case produce a very efficient braking \citep{Meynet2011a}, which would make the formation of a collapsar nearly impossible\footnote{It would be extremely interesting in that respect to conduct spectropolarimetric observations to measure the geometry and the strength of the surface magnetic field, coupled with asteroseismological observations giving information on the internal variation of the angular velocity. This would, for instance, answer questions such as whether the stars presenting a strong surface magnetic field rotate as solid body in the interior.}.

From the points listed above, we are inclined to think that most of the massive stars do not rotate as solid bodies during their MS phase but possess a differentially rotating interior. If true, this implies that the slowing-down of the core needed to reproduce the observed periods of young pulsars should result from another process than the strong coupling caused by an internal magnetic field. Could it be caused by a surface magnetic braking? A sufficiently strong surface magnetic field may indeed force the wind material to follow the magnetic lines. Conservation of angular momentum will exert a torque and thus slow down the star \citep{ud-Doula2009a}\footnote{The magnetic field could also drive the matter back to the surface of the star and thus prevent any loss of angular momentum. In that case, the magnetic field would not help as a braking mechanism.}. It does appear now that at least some massive stars show a surface magnetic field \citep[see the review by][]{Petit2011a}.  We show in Fig.~\ref{BminRSG} the minimum value of a dipole-aligned surface magnetic field needed to interact with the stellar winds. We see that on the MS, for stars with masses between $20$ and $40\,M_{\sun}$, even a $<200\, \text{G}$ magnetic field would be sufficient to impose some coupling between the wind and the surface and would therefore slow the star down. The minimum magnetic field allowing a coupling with the wind is even lower during the RSG phase, with a couple of $10\, \text{G}$. In stars more massive than about $50\, M_{\sun}$, magnetic fields between $300$ and $600\,\text{G}$ are needed during the MS phase, and are of the order of a few $10000\,\text{G}$ during the WR phase. The consequences of such a surface magnetic field braking mechanism remain to be studied  \citep[see first results in][]{Meynet2011a}.

We finally note that a large difference between models with solid-body rotation (and thus ``strong''  internal coupling) and those with internal differential rotation (``moderate'' internal coupling), such as those presented here, is the manner in which LGRB occurrence may depend on metallicity. For strong coupling, the occurrence of LGRB is mostly restricted  to low metallicity where the mass loss by stellar winds is weak. For moderate coupling a metallicity dependence  is also present, but LGRB could occur at solar metallicity, as illustrated by the present models,  and probably at higher than solar metallicity \citep{Levesque2010d,Levesque2010a}. Very interestingly, the lack of correlation between the energy of the explosion of LGRB with metallicity as obtained by \citet{Levesque2010c} may reflect the fact that the internal conditions, which govern the energy, are not strongly impacted by metallicity. This somehow supports the ``moderate'' coupling scenario, in which the properties of the core, in terms of angular momentum content, are much less sensitive to metallicity than in the ``strong''  coupling scenario.


\section{Synthesis and conclusion \label{SecConclu}}

We have presented predictions of single-star models and compared them to several important observational features: the populations of WR stars, of type Ibc SNe, and of collapsars at solar metallicity. 

We confirm with the new generation of rotating models and the results obtained from previous work \citep{Meynet2003a,Meynet2005a,Georgy2009a} that rotation favours the formation of WR and type Ibc SNe.

On the basis of these new models, we estimate that the single-star channel could account for  about 60\% of the observed WR stars at solar metallicity, with the remaining 40\% likely originating from close-binary systems.

The fraction of type Ibc SNe coming from the single-star channel will remain difficult to obtain in a reliable way until SN models are able to reliably predict the properties of the explosion of core-collapse events leading to BH formation.  At the moment, all we can say is that at least $\sim$50\% of type Ibc SNe could originate from the single-star scenario.

The present rotating models predict a lack of type II-P progenitors above $\sim 17\, M_{\sun}$ while the non-rotating ones predict such a lack only above $19\, M_{\sun}$. Both models therefore agree with the observations by \citet{Smartt2009a}, who observed no type II-P SN progenitors with initial masses above about $18\, M_{\sun}$. Moreover, rotating models provide an excellent explanation for the characteristics of the progenitor of the SN 2008ax.

We also discuss some possible shortcomings of the present models. Part of these shortcomings may also be due to uncertainties in the observations, or to a presently biased view linked to incomplete sampling. Based on the available observations, it does appear that the current rotating models may overestimate the duration of the WNL phase and underestimate those of the WNE and WC phase. A possible reason for this is the fact that the present recipe for the WNL mass-loss rate predicts very low values \citep{Grafener2008a}. 

Concerning the origin of the low-luminosity WC stars, the results may be in part improved by an enhancement of the mass-loss rates during the WNL phase. An enhancement of the mass-loss rates during the RSG phase and/or mass transfer in close-binary systems could also help resolve this problem.

The question of the rotation rate of pulsars indicates some additional angular-momentum loss from the central region with respect to the scenario obtained in our models. While a strong internal coupling (caused, for instance, by an internal magnetic field) might resolve this problem, we mentioned alternative solutions. In particular, the role played by magnetic braking at the surface needs to be explored in more detail in this context. 

From a theoretical point of view, the conditions for having collapsars are not necessarily restricted to low metallicity. The effect of metallicity on the evolution leading to a LGRB is strongly correlated to the degree of coupling between the core and the envelope in stellar models. A strong coupling is more prone to confine GRB occurrence to low metallicity than a moderate or a weak coupling.

\begin{acknowledgements}
R. H. acknowledges support from the World Premier International Research Center Initiative (WPI Initiative), MEXT, Japan. We also thanks Andreas Sander for
communicating us observational data in press as well as Wolf-Rainer Hamann for his very useful referee report, and J. J. Eldridge for his helpful comments.
\end{acknowledgements}

\bibliographystyle{aa}
\bibliography{MyBiblio}

\end{document}